\documentclass[a4paper,11pt]{article}
\pdfoutput=1 % if your are submitting a pdflatex (i.e. if you have
\usepackage{jcappub} % for details on the use of the package, please
\usepackage[T1]{fontenc} % if needed
\usepackage[utf8]{inputenc} %for a bib tex reference with special character

\newcommand{\newform}[2]{#1}

\title{Can power spectrum observations rule out slow-roll inflation?}

\author[a]{J. P. P. Vieira,}
\author[a]{Christian T. Byrnes,}
\author[a]{and Antony Lewis}

\affiliation[a]{Department of Physics \& Astronomy, University of Sussex, Brighton BN1 9QH, UK}

\emailAdd{J.Pinto-Vieira@sussex.ac.uk}

\abstract{The spectral index of scalar perturbations is an important observable that allows us to learn about inflationary physics. In particular, a detection of a significant deviation from a constant spectral index could enable us to rule out
the simplest class of inflation models.
We investigate whether future observations could rule out canonical single-field slow-roll inflation given the parameters allowed
by current observational constraints. We find that future measurements of a constant running (or running of the running)
of the spectral index over currently available scales are unlikely to achieve this.
However, there remains a large region of parameter space (especially when considering the running of the running)
for falsifying the assumed class of slow-roll models if future observations accurately constrain a much
wider range of scales.}
\begin{document}
\maketitle
\flushbottom

%\tableofcontents

\section{Introduction}

One of the main achievements of the recent era of precision cosmology
has been the increasing quality of measurements of the cosmic microwave
background (CMB) across the sky, for example by the
\emph{Planck} mission \cite{Planck_inf2015}. These have been invaluable
in constraining physics in the very early Universe. In particular,
these measurements can be used to measure the scale-dependence of the primordial power
spectrum, and have been instrumental in establishing
cosmic inflation as the most popular paradigm for the universe before the hot big bang.

Despite this success, so far only two perturbation parameters of relevance to inflationary models have been measured to be non-zero:
the amplitude of the scalar power spectrum and its spectral index, $n_{s}$. One consequence of this lack of measured
observables is a difficulty in differentiating between different specific models of inflation,
though the non-observation of primordial tensor modes already provides a powerful constraint on broad classes of inflationary models \cite{Martin:2013tda}.
%which is necessary for learning about the subsequent epoch of reheating
%as well as generally about particle physics at very high energies.
%As such,
Finding new measurable observables that could falsify some of the remaining allowed models
is one of the main goals of modern cosmology.

Although recent attempts at finding such observables have focused mostly on non-Gaussian signals
in higher-order correlation functions \cite{Vennin:2015egh,Renaux-Petel:2015bja},
there are still a few relevant quantities at the level of the power spectrum whose precision should be noticeably
improved by future probes \cite{Adshead:2010mc,Kohri:2013mxa,mudistorrunning,Munoz:2016owz,Sekiguchi:2017cdy}.
The running ($\alpha_{s}$) and the running of the
running ($\beta_{s}$) of the spectral index of scalar perturbations are examples of parameters that can be measured
more accurately in the future and are predicted to have very small magnitude
(compared to $n_{s}-1$) in the simplest classes of canonical single-field slow-roll inflation.
This is especially interesting because, even though
current constraints on these quantities are compatible with zero,
their best-fit values have an amplitude comparable to $n_{s}-1$ \cite{BKPanalysis,Planck_inf2015,running_running}. A future detection of $\alpha_s$ or $\beta_s$ could in principle
provide strong evidence against these simplest classes of inflationary models.

While a detection of $\alpha_{s}$ or $\beta_{s}$ at the same order as $n_s-1$ would rule out the simplest slow-roll models, the implications for the wider class of canonical single-field slow-roll inflation models
are less obvious and require a more general treatment.
In this paper, we study the more general implications using the well-studied formalism
for computing power spectra developed in Refs.~\cite{STEWART_RUN,DOD_STEWART,Choe_Gong_Stewart_second,Stewart_inverse1,Stewart_inverse2}.
Although we fall short of a completely generic conclusion,
our results are sufficient to show that it is much harder to rule out slow roll than the simplest arguments suggest.

Section \ref{sec:Generalised-Slow-Roll} of this paper is devoted
to motivating our treatment and introducing the formalism it is based on; section \ref{sec:Exploring-the-limits-of-slow-roll}
explains how to assess whether specific values of $\alpha_{s}$ and $\beta_{s}$ are compatible with
slow-roll inflation; and section \ref{sec:Results} presents the
results (with the main technical details of the calculations being left to the
appendices), including a comparison with current observational
bounds (effectively extending the analysis made with WMAP data in \cite{Easther:2006tv}). Finally, in section
\ref{sec:Conclusions}, we summarize our conclusions, including
a discussion of future prospects.

Throughout this work we assume a $\Lambda$CDM cosmology evolving according to general relativity seeded by fluctuations from single-field inflation, and use natural units with $c=\hbar=M_{P}^{2}=\left(8\pi G\right)^{-1}=1$.

\section{General slow-roll approximation\label{sec:Generalised-Slow-Roll}}

In canonical single-field inflation, the energy density of the Universe
is dominated by that of a scalar field, $\phi$ (the inflaton), and
thus the Hubble parameter of a flat FLRW metric is given by the first
Friedmann equation as
\begin{equation}
3H^{2}=\frac{1}{2}\dot{\phi}^{2}+V\left(\phi\right),\label{eq:Friedmann1}
\end{equation}
where $V$ is the inflaton potential and $H$ is the Hubble parameter.
The inflaton obeys the equation of motion
\begin{equation}
\ddot{\phi}+3H\dot{\phi}+V^{\prime}\left(\phi\right)=0,\label{eq:inflaton_eom}
\end{equation}
where the prime denotes differentiation with respect to argument (here with respect to $\phi$)
and the dot denotes differentiation with respect to time.

A simplifying assumption often used to study inflation models
is the slow-roll approximation, which states that the inflaton rolls
down its potential slowly enough that:
\begin{enumerate}
\item its kinetic energy is much less than its potential energy, i.e.,
\begin{equation}
\epsilon\equiv-\frac{\dot{H}}{H^{2}}=\frac{1}{2}\left(\frac{\dot{\phi}}{H}\right)^{2}\ll1;\label{eq:eps_small}
\end{equation}

\item $\ddot{\phi}$ can be neglected in Eq.~\eqref{eq:inflaton_eom}, i.e.,
\begin{equation}
\left|\delta_{1}\right|\equiv\left|\frac{\ddot{\phi}}{H\dot{\phi}}\right|\ll1.\label{eq:dealta1_small}
\end{equation}

\end{enumerate}
If this simplification is valid (which is the case for most models compatible with observations),
it is straightforward to compute the evolution of background quantities from the slow-roll equations
\begin{equation}
3H^{2}\simeq V,\label{eq:Friedmann1-1}
\end{equation}
\begin{equation}
3H\dot{\phi}+V^{\prime}\simeq0\label{eq:inflaton_eom-1}
\end{equation}
(which follow trivially from applying the slow-roll approximation to
Eqs. \eqref{eq:Friedmann1} and \eqref{eq:inflaton_eom}, respectively).
%\AL{You've said "usually sufficient" to do this.. you meant to follow up with some limitation here?}
%\JV{Not exactly. I wasn't referring specifically to "models compatible with observations" because at this point
%we haven't discussed observable stuff yet. I meant "usually" in the context of all %models (where I agree the word has a less than
%well-defined meaning).}

The quantities $\epsilon$ and $\delta_{1}$ defined above are known as the slow-roll parameters
(note that there are several popular alternative definitions and notations for $\delta_{1}$).
It is also possible to define ``higher-order'' slow-roll parameters, for example as
\begin{equation}
\delta_{n}\equiv\frac{1}{H^{n}\dot{\phi}}\frac{d^{n}\dot{\phi}}{dt^{n}}.\label{eq:delta_ndef}
\end{equation}
Although these parameters are not strictly important for establishing whether the
slow-roll approximation is valid, in practice it is often necessary to make assumptions regarding their relative smallness
in order to be able to compute the corresponding spectrum of scalar perturbations consistently to a given order.

\subsection{The scalar power spectrum in slow-roll inflation}

As previously noted by Stewart and Gong \cite{STEWART_RUN,STEWART_GONG},
the slow-roll approximation is not always sufficient
to accurately calculate the power spectrum of scalar perturbations.

The equation of motion for the Fourier modes of the scalar perturbations
is \cite{SL1993}
\begin{equation}
\frac{d^{2}\varphi_{k}}{d\xi^{2}}+\left(k^{2}-\frac{1}{z}\frac{d^{2}z}{d\xi^{2}}\right)\varphi_{k}=0,\label{eq:fourier_eom}
\end{equation}
where $z\equiv\frac{a\dot{\phi}}{H}$, the gauge-invariant curvature perturbation is $-\varphi_k / z$,
$\xi\equiv -\eta$ is minus the conformal time (varying from $\infty$
in the infinite past to $0$ in the infinite future),
and we assume asymptotic boundary conditions
\begin{equation}
\varphi_{k}\longrightarrow\begin{cases}
\frac{e^{ik\xi}}{\sqrt{2k}}, & k\xi\rightarrow\infty\\
A_{k}z, & k\xi\rightarrow0
\end{cases},\label{eq:fourier_boundary}
\end{equation}
 where $A_{k}$ is a constant for each wave vector $k$.

To keep track of the approximations
that will be needed, it is useful to use the rescaled variables
\begin{equation}
y\equiv\sqrt{2k}\varphi_{k},\label{eq:ydef}
\end{equation}
\begin{equation}
x\equiv k\xi.\label{eq:xdef}
\end{equation}
Using these we can  rewrite the equation of motion for each Fourier mode as
\begin{equation}
\frac{d^{2}y}{dx^{2}}+\left(1-\frac{2}{x^{2}}\right)y=\frac{g\left(\ln x\right)}{x^{2}}y,\label{eq:y_eom}
\end{equation}
where the important function $g$ is defined in terms of
\begin{equation}
f\left(\ln\xi\right)\equiv\frac{2\pi a\xi\dot{\phi}}{H}\label{eq:fdef}
\end{equation}
as
\begin{equation}
g\left(\ln x\right)\equiv\left[\frac{f^{\prime\prime}-3f^{\prime}}{f}\right]_{\xi=\frac{x}{k}}.\label{eq:gdef}
\end{equation}

The power spectrum can be straightforwardly (although not necessarily easily) calculated by solving
Eq.~\eqref{eq:y_eom} and then finding
\begin{equation}
\mathcal{P}\left(k\right)=\lim_{x\rightarrow0}\left|\frac{xy}{f}\right|^{2}.\label{eq:Ps_y}
\end{equation}
The homogeneous solution (for $g=0$),
\begin{equation}
y_{0}\left(x\right)=\left(1+\frac{i}{x}\right)e^{ix},\label{eq:y0}
\end{equation}
together with the relation (which is justified later in appendix \ref{sec:gappendix})
\begin{equation}
\xi=\frac{1}{aH}\left(1+\mathcal{O}\left(g\right)\right)\label{eq:eta_aH}
\end{equation}
lead, at zeroth order in $g$, to the simple scale-invariant\footnote{It can be seen
(for example through Eq.~\eqref{eq:Friedmann1}) that this result is divergent, as expected in a de Sitter background. This is not a problem
as all that matters is that when this becomes the leading contribution to a more
realistic power spectrum it is approximately scale-invariant (which is guaranteed by the slow-roll approximation).}
power spectrum
\begin{equation}
\mathcal{P}_{0}\left(k\right)=\lim_{x\rightarrow0}\left|\frac{i}{f}\right|^{2}=\frac{H^{4}}{\left(2\pi\dot{\phi}\right)^{2}}.\label{eq:standard_Ps}
\end{equation}

The standard slow-roll result can then be obtained by arguing that
in a more general slow-roll scenario (with small $g\neq0$) the leading
contribution to the power spectrum (with corrections being suppressed
by terms of order $g$) will still be given by Eq.~\eqref{eq:standard_Ps}
if the now non-constant terms are evaluated at some point around horizon crossing.

\subsection{The spectral index in general slow-roll inflation\label{specindex_GSRS}}

The slow-roll approximation has been sufficient
to derive the standard lowest-order result of Eq.~\eqref{eq:standard_Ps}.
However, to derive the standard first-order prediction for the spectral index \cite{Liddle:1992wi},
\begin{equation}
n_{s}-1\equiv\frac{d\ln\mathcal{P}}{d\ln k}=-4\epsilon-2\delta_{1}\label{eq:ns_1st_standard}
\end{equation}
(where the slow-roll parameters are to be evaluated around the time of horizon crossing),
 the first-order corrections to Eq.~\eqref{eq:standard_Ps}
must only give at most a second-order contribution to $n_{s}-1$,
which is not true in general. Ignoring those corrections, as is usually done, requires a hierarchy for higher-order slow-roll parameters such that~\cite{STEWART_RUN,DOD_STEWART}
\begin{equation}
\left|\delta_{n+1}\right|\ll\left|\delta_{n}\right|,\label{eq:usualhierarchy}
\end{equation}
which does not necessarily follow
from the ``vanilla'' slow-roll assumptions.

Assuming this hierarchy of slow-roll parameters, the leading-order prediction for the running of the spectral index becomes
\begin{equation}
\alpha_{s}\equiv\frac{dn_{s}}{d\ln k}=-2\delta_{2}-8\epsilon^{2}+2\delta_{1}^{2}-10\epsilon\delta_{1},\label{eq:alpha_def}
\end{equation}
so that (barring fine-tuning effects) $\alpha_{s} \sim {\cal{O}} \left(\left| n_{s}-1\right|^{2}\right)\ll\left| n_{s}-1\right|$,
which motivates the naive expectation that $\alpha_{s}$ be negligible in slow-roll inflation. Mutatis mutandis, it can be seen that the equivalent expectation for the running of the running,
\begin{equation}
\beta_{s}\equiv\frac{d^{2}n_{s}}{d\ln k^{2}},\label{eq:beta_def}
\end{equation}
is that $\left|\beta_{s}\right|\sim{\cal{O}}\left(\left| n_{s}-1\right|^{3}\right)\ll\left| \alpha_{s}\right|\ll\left|n_{s}-1\right|$.

\begin{figure}[h]

\includegraphics[scale=0.95]{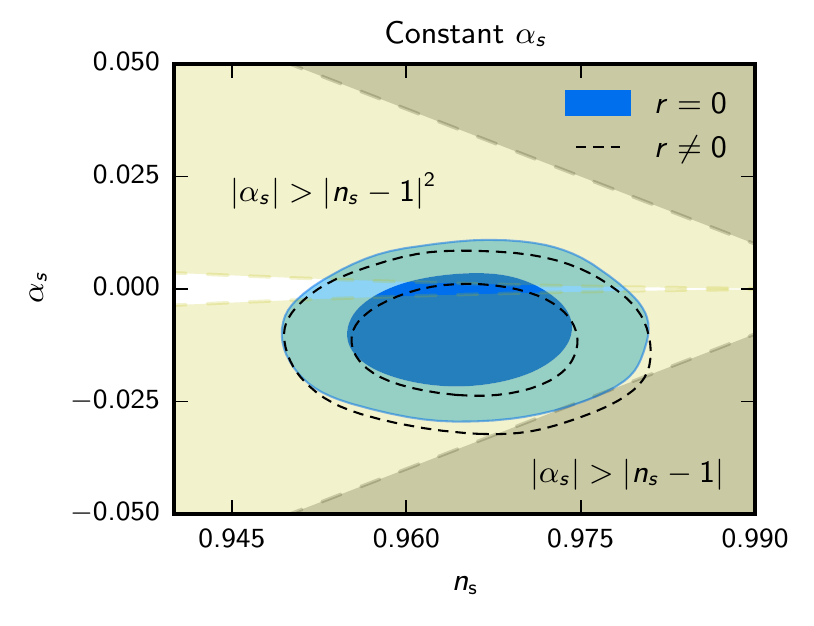}
\includegraphics[scale=0.95]{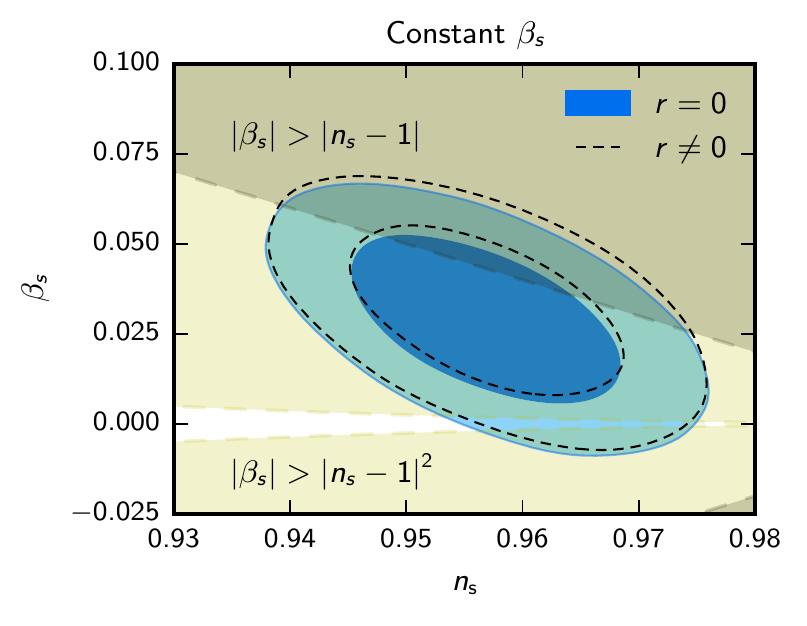}

\caption{\label{fig:motivation} Constraints  from \emph{Planck} 2015 TT+lowTEB~\cite{Planck_inf2015} and BICEP-Keck~\cite{Array:2015xqh} on a constant $\alpha_{s}$ (left) and
a constant $\beta_{s}$ (marginalized over $\alpha_{s}$ at the pivot scale; right), both against $n_{s}$ at the pivot scale.
Dashed black contours assume a null tensor-to-scalar ratio, $r$, whereas blue contours marginalize over it.
The light shaded region corresponds to the part of the parameter space where the quantity in the vertical axis becomes greater than
$\left|n_{s}-1\right|^{2}$ and the dark shaded region is where it becomes greater
than $\left|n_{s}-1\right|$. The naive expectation is that the true value of $\alpha_{s}$ (left) should be close to the boundary of the
unshaded region and far away from the dark shaded region, whereas that of $\beta_{s}$ (right) should be well inside the
unshaded region. Current constraints allow a much greater area of the parameter space.}

\end{figure}

As is shown in figure \ref{fig:motivation}, although current constraints are consistent with small $\alpha_{s}$ and $\beta_{s}$ as predicted by the naive hierarchy, much larger values are still currently allowed. Indeed, the
posteriors currently peak substantially away from zero, especially for $\beta_{s}$ (largely due to the low-$\ell$ feature in the CMB temperature power spectrum \cite{Ade:2013zuv}).
Improved future constraints\footnote{Note that next-generation
missions may improve these bounds by about an order of magnitude \cite{Munoz:2016owz,Kohri:2013mxa}.} on $\alpha_{s}$ and $\beta_{s}$ that peak away from zero could rule out
the simplest class of inflationary models (characterized by the slow-roll approximation
and the hierarchy in Eq.~\eqref{eq:usualhierarchy}), but a more general statement about
the wider class of canonical single-field slow-roll inflation models requires a more general treatment.
%\JV{Is this still OK? I feel that emphasising the great freedom we currently have to go beyond this
%expectation is worth emphasising a little...}
%\AL{I've split the super-long sentence up again!}

A few ways to approach modelling a more general slow-roll scenario are available in the literature \cite{STEWART_RUN,Choe_Gong_Stewart_second,DVORKIN_GSR,Stewart_inverse2,
Stewart_inverse1,GSR_inflEFT,GSRfeatures,GSRtensor,GSRinin,Adshead:2013zfa}.
In this work, we use the results from Ref.~\cite{Stewart_inverse2}
(which in turn use the results from Ref.~\cite{Choe_Gong_Stewart_second}), which we briefly review.

To solve for the power spectrum (Eq.~\eqref{eq:Ps_y}) we need to solve for the modes $y$.
The second-order linear differential equation in Eq.~\eqref{eq:y_eom} can be solved for $y$ using Green's functions, with solution satisfying the boundary conditions of Eq.~\eqref{eq:fourier_boundary} given implicitly by
\begin{equation}
y\left(x\right)=y_{0}\left(x\right)+\frac{i}{2}\intop_{x}^{\infty}\frac{du}{u^{2}}g\left(\ln u\right)\left[y_{0}^{*}\left(u\right)y_{0}\left(x\right)-y_{0}^{*}\left(x\right)y_{0}\left(u\right)\right]y\left(u\right).\label{eq:integral_y}
\end{equation}
This can be solved iteratively for $y$ to successively higher order in $g$ (assuming $\left|g\right|<1$) by substituting
the previous order result into the right-hand
side of Eq.~\eqref{eq:integral_y} (starting with $y\left(u\right)=y_{0}\left(u\right))$.
The result for the power spectrum at the desired order can then be obtained by substituting into Eq.~\eqref{eq:Ps_y}
and simplifying as much as possible. The result for the scalar power spectrum correct to quadratic order in $g$ is then \cite{Choe_Gong_Stewart_second}
\begin{multline}
\ln\mathcal{P}\left(\ln k\right)=\intop_{0}^{\infty}\frac{d\xi}{\xi}\left[-k\xi W^{\prime}\left(k\xi\right)\right]\left[\ln\frac{1}{f\left(\ln\xi\right)^{2}}+\frac{2}{3}\frac{f^{\prime}\left(\ln\xi\right)}{f\left(\ln\xi\right)}\right]+\frac{\pi^{2}}{2}\left[\intop_{0}^{\infty}\frac{d\xi}{\xi}m\left(k\xi\right)\frac{f^{\prime}\left(\ln\xi\right)}{f\left(\ln\xi\right)}\right]^{2}
\\
-2\pi\intop_{0}^{\infty}\frac{d\xi}{\xi}m\left(k\xi\right)\frac{f^{\prime}\left(\ln\xi\right)}{f\left(\ln\xi\right)}\intop_{\xi}^{\infty}\frac{d\zeta}{\zeta}\frac{1}{k\zeta}\frac{f^{\prime}\left(\ln\zeta\right)}{f\left(\ln\zeta\right)}+{\cal{O}}\left(g^{3}\right),\label{eq:P_f_2nd}
\end{multline}
where $W$ and $m$ are window functions defined by
\begin{equation}
W\left(x\right)=\frac{3\sin\left(2x\right)}{x^{3}}-\frac{3\cos\left(2x\right)}{x^{2}}-\frac{3\sin\left(2x\right)}{2x}-1\label{eq:Wdef}
\end{equation}
 and
\begin{equation}
m\left(x\right)=\frac{2}{\pi}\left[\frac{1}{x}-\frac{\cos\left(2x\right)}{x}-\sin\left(2x\right)\right].\label{eq:mdef}
\end{equation}

In this paper we are interested in relating properties of the observable power spectrum to those of the inflationary model,
so we need the inverse version of this result, which can be shown to be \cite{Stewart_inverse2}
\newform{
\begin{multline}
\ln\frac{1}{f\left(\ln\xi\right)^{2}}= \intop_{0}^{\infty}\frac{dk}{k}m\left(k\xi\right)\ln\mathcal{P}\left(\ln k\right)
-\frac{\pi^2}{8}\intop_{0}^{\infty}\frac{dk}{k}m\left(k\xi\right)
\left[\intop_{0}^{\infty}\frac{dl}{l}
 \frac{\mathcal{P}'(\ln l)}{\mathcal{P}(\ln l)}
 \intop_{0}^{\infty}\frac{d\zeta}{\zeta}m\left(k\zeta\right)m\left(l\zeta\right)
 \right]^{2}
\\+\frac{\pi}{2}\intop_{0}^{\infty}\frac{dl}{l} \frac{\mathcal{P}'(\ln l)}{\mathcal{P}(\ln l)} \intop_{0}^{\infty}\frac{dq}{q}
\frac{\mathcal{P}'(\ln q)}{\mathcal{P}(\ln q)}
\intop_{0}^{\infty}\frac{d\zeta}{\zeta}m\left(l\zeta\right)\intop_{0}^{\infty}\frac{dk}{k^{2}}m\left(k\xi\right)m\left(k\zeta\right)\intop_{\zeta}^{\infty}\frac{d\chi}{\chi^{2}}m\left(q\chi\right).
\label{eq:inv2_formula}
\end{multline}
}{
\begin{multline}
\ln\frac{1}{f\left(\ln\xi\right)^{2}}=\intop_{0}^{\infty}\frac{dk}{k}m\left(k\xi\right)\ln\mathcal{P}\left(\ln k\right)
\\
-\frac{1}{2\pi^{2}}\intop_{0}^{\infty}\frac{dk}{k}m\left(k\xi\right)\left[\intop_{0}^{\infty}\frac{dl}{l}\ln\left|\frac{k+l}{k-l}\right|\frac{\mathcal{P}^{\prime}\left(\ln l\right)}{\mathcal{P}\left(\ln l\right)}\right]^{2}
\\
+\intop_{0}^{\infty}\frac{dl}{l}\intop_{0}^{\infty}\frac{dq}{q}M\left(l\xi,q\xi\right)\frac{\mathcal{P}^{\prime}\left(\ln l\right)}{\mathcal{P}\left(\ln l\right)}\frac{\mathcal{P}^{\prime}\left(\ln q\right)}{\mathcal{P}\left(\ln q\right)},\label{eq:inv2_formula}
\end{multline}
where we have introduced the new window function $M$, which can be
written as
\begin{equation}
M\left(x,y\right)=\frac{2}{\pi^{2}xy}\left[h\left(x\right)+h\left(y\right)-\frac{1}{2}h\left(x-y\right)-\frac{1}{2}h\left(x+y\right)\right],\label{eq:Mdef}
\end{equation}
where we have defined
\begin{equation}
h\left(x\right)=x\,\mathrm{Si}\left(2x\right)+\frac{\cos\left(2x\right)}{2}-\frac{1}{2},\label{eq:h_g_M}
\end{equation}
where $\mathrm{Si}$ stands for the sine integral function,
\begin{equation}
\mathrm{Si}\left(x\right)\equiv\intop_{0}^{x}\frac{\sin t}{t}dt.\label{eq:sinint}
\end{equation}
}

\section{Exploring the limits of slow-roll\label{sec:Exploring-the-limits-of-slow-roll}}

\subsection{How slow is slow-roll?}

To assess how much running there can be in slow-roll inflation, we would like
some objective criteria to decide whether any given
inflationary model is slow-roll or not.
The ``$\ll$'' signs in Eqs. \eqref{eq:eps_small}-\eqref{eq:dealta1_small}
defining the slow-roll approximations do not allow a clear distinction unless the numbers being compared are
orders of magnitude apart. To make matters worse, Eq.~\eqref{eq:dealta1_small}
has been defined in the literature in terms of a number of slightly
different slow-roll parameters (usually referred to as $\eta$), all
of which would lead to different classifications of borderline
cases even if we were to decide on an objective meaning for ``$\ll$''
in these equations.

When faced with this sort of problem it is important not to get lost
in an overly semantic discussion. One pragmatic reason to care about whether a model falls under the
category of slow-roll is simply to know whether the power spectrum can be straightforwardly computed
using results like Eq.~\eqref{eq:standard_Ps} and Eq.~\eqref{eq:P_f_2nd}.
Therefore, from the perspective of this work, the best way to define
slow-roll is in terms of a quantity that can quantify how precise
this formula actually is. From the derivation, the
most natural quantity appears to be the parameter $g$. Unfortunately, this will result in a slightly stronger definition than using
just the slow-roll approximation, as it discards scenarios in which $\delta_{2}$ is large
but $\epsilon$ and $\delta_{1}$ remain small (see appendix \ref{sec:gappendix}).
Nevertheless, it is a weak enough definition that we will be able to
qualitatively improve on the simplistic constraints in subsection \ref{specindex_GSRS}\footnote{
To calculate the power spectra for specific slow-roll potentials,
one could always resort to the more general formalism of Generalized Slow-Roll \cite{DVORKIN_GSR},
which relies on a weaker assumption than the slow-roll approximation (allowing for even $\delta_{1}$ to
become large for short periods of time). However, our analysis would be much more complicated
in that context, both due to difficulties in defining slow-roll (which is the regime we are interested in here)
and due to the added difficulties in solving the inverse problem of finding the model that corresponds
to a given power spectrum.}.

Instead of committing to any arbitrary definition of what a ``very small'' number is, we show,
for each combination of observable parameter values, how large $g$ can become during the period of time
in which observable scales crossed the horizon. The reader can not only decide which values are
``not small'' on his/her own, but also have a good understanding of the meaning of any specific choice:
the larger the allowed values, the less accurate our formulas.

\subsection{Outline of the method}

%We now have all the tools to do what we set out to do at the beginning.
%In order to do this,
We start by parameterizing the observed scalar power spectrum as
\begin{equation}
\ln\mathcal{P}\left(\ln k\right)=\sum_{n=0}^{N}\frac{\beta_{n}}{n!}\left(\ln\frac{k}{k_{0}}\right)^{n},\label{eq:P_ansatz}
\end{equation}
where $k_{0}$ is a pivot scale and the $\beta_{n}$ coefficients
are to be constrained by observations. Of course,
\begin{equation}
\begin{split}
& \beta_{0}\equiv\ln\mathcal{P}_{0} \\
& \beta_{1}\equiv n_{s}-1 \\
& \beta_{2}\equiv\alpha_{s} \\
& \beta_{3}\equiv\beta_{s}
\label{eq:betans}
\end{split}
\end{equation}
where $\mathcal{P}_{0}$ is the magnitude of the power spectrum at
the pivot scale. For the purposes of this work, we will be interested
in the cases with $N=2$ and $N=3$, for which $\beta_{N}$ have already
been constrained by the \emph{Planck} collaboration \cite{Planck_inf2015}\footnote{
Other works \cite{running_running} have claimed slightly more dramatic constraints for the
$N=3$ case.}. A natural extension of our calculations is sufficient to deal with cases with higher $N$ should observational constraints
on higher-order runnings become available (it has been claimed such constraints could come from minihalo effects on 21cm fluctuations \cite{Sekiguchi:2017cdy}). Likewise, radically different parameterizations of the power spectrum
can be incorporated by making the appropriate changes to Eq.~\eqref{eq:P_ansatz}.

For each point in the $\left(\beta_{0},\beta_{1},...,\beta_{N}\right)$
parameter space, we want to know to what extent a canonical single-field inflation model
must violate slow-roll during the interval of time during which observable scales left the horizon
(i.e., how large its respective $g$ function must become during that time).
%in order to be able to reproduce the observed power spectrum that corresponds to it.
%With this in mind,

We proceed by defining a $g\left(\ln x\right)$ for every $k$ by inserting the
power spectrum from Eq.~\eqref{eq:P_ansatz}\footnote{We can ignore the term
with $\beta_{0}$ since, from Eq.~\eqref{eq:inv2_formula},
it only contributes to a proportionality constant in $f$, and thus
has no effect on $g$.%
} into Eq.~\eqref{eq:inv2_formula}, and then the resulting
$f\left(\ln\xi\right)$ into Eq.~\eqref{eq:gdef}.
The main obstacle in the way of this calculation is the computation
of the integrals in Eq.~\eqref{eq:inv2_formula}
when the power spectrum is a polynomial in $\ln \frac{k}{k_{0}}$, as we assume (in Eq.~\eqref{eq:P_ansatz}).
%Note that we can be assured
%that the rest of this calculation method will be trivial once these
%have been found, for it can be seen by a change of variables of integration
%that the right-hand side of Eq.~\eqref{eq:inv2_formula} must be a polynomial
%in $\ln\left(k_{0}\xi\right)$ - and thus easily differentiable.
To solve these integrals for power spectra with non-null $\beta_{s}$, we extended
known results for the standard hierarchy in the slow-roll approximation \cite{Stewart_inverse2} (see
appendix \ref{sec:Solving-the-integrals}).
The results are polynomials in $\ln\left(k_{0}\xi\right)$, so $g$ can then be found straightforwardly
from Eq.~\eqref{eq:gdef} by differentiation.

Once these $g$ functions have been found for $k$ corresponding to observable scales, we check the absolute value of $g$
at $x=1+\epsilon=1+\frac{r}{16}$, corresponding to the time of horizon-crossing to leading order in $\epsilon$
(see, e.g., Eqs.~\eqref{eq:eta_epsbar} and \eqref{eq:eps_bar_slow_final} in appendix \ref{sec:gappendix})\footnote{
The reason we are justified in resorting to a first-order result after using second-order results up until this point
is that $r$ is already observationally constrained to be so small that even the leading order term has
no significant effect on our constraints.}.

\section{Results\label{sec:Results}}

The method described in section \ref{sec:Exploring-the-limits-of-slow-roll}
was implemented in a Python code using the results from appendix \ref{sec:Solving-the-integrals}.
This allowed us to draw contour plots indicating how large $g$ can
get during the relevant epochs for different pivot values of $n_{s}$,
$\alpha_{s}$, and $\beta_{s}$, assuming that Eq.~\eqref{eq:P_ansatz}
holds for a specific range of observable scales. In this work we present results
for three different ranges of observable scales, from $k_{\rm min}=10^{-3}\mathrm{Mpc^{-1}}$ (set by the largest scales that can be
reasonably well measured) up to: $k_{\rm max}=0.3\mathrm{Mpc^{-1}}$ (spanning about 6 efoldings), roughly
corresponding to the smallest scale well constrained by Planck;
$k_{\rm max}=100\mathrm{Mpc^{-1}}$ (spanning about 12 efoldings), roughly corresponding
to a future constraint from 21cm observations \cite{Loeb:2003ya,Sekiguchi:2017cdy}; and $k_{\rm max}=10^4\mathrm{Mpc^{-1}}$
(spanning about 16 efoldings), roughly corresponding to the smallest scale constrained by spectral distortions \cite{Chluba:2012we}\footnote{We note that the supernova lensing dispersion can also probe the averaged value of the power spectrum on small scales, down to $k_{\rm max}\gtrsim 100 \mathrm{Mpc^{-1}}$, but there is a degeneracy with the effect of baryons on the small-scale low-redshift power spectrum \cite{Ben-Dayan:2015zha}.}.
The pivot scale is taken to be $k_{0}=0.05\mathrm{Mpc^{-1}}$, the \emph{Planck} pivot scale\footnote{Note
that this value is only important for including observational constraints in our plots. Naturally, if one merely wanted to know
how large a constant $\alpha_{s}$ or $\beta_{s}$ are allowed to be over a certain range of scales,
the pivot scale would be irrelevant (for example because it plays no role in Eq.~\eqref{eq:inv2_formula}).}.

In order to make statements about the status of this class of canonical single-field slow-roll inflation,
we use CosmoMC \cite{Lewis:2002ah,Lewis:2013hha} to superimpose
current constraints from \emph{Planck} 2015 data (temperature plus low-$\ell$ polarization, TT + lowTEB~\cite{Planck_inf2015})
and the latest BICEP-KECK-\emph{Planck} joint analysis~\cite{Array:2015xqh}, showing the
 $1\sigma$ and $2\sigma$ allowed regions.
Additionally, we plot contours for the inferred maximum values of $g$ from current observations
against $\alpha_{s}$ and $\beta_{s}$.

\subsection{Constant running ($N=2$)}

\begin{figure}[h]
\begin{center}
\includegraphics[scale=1.0]{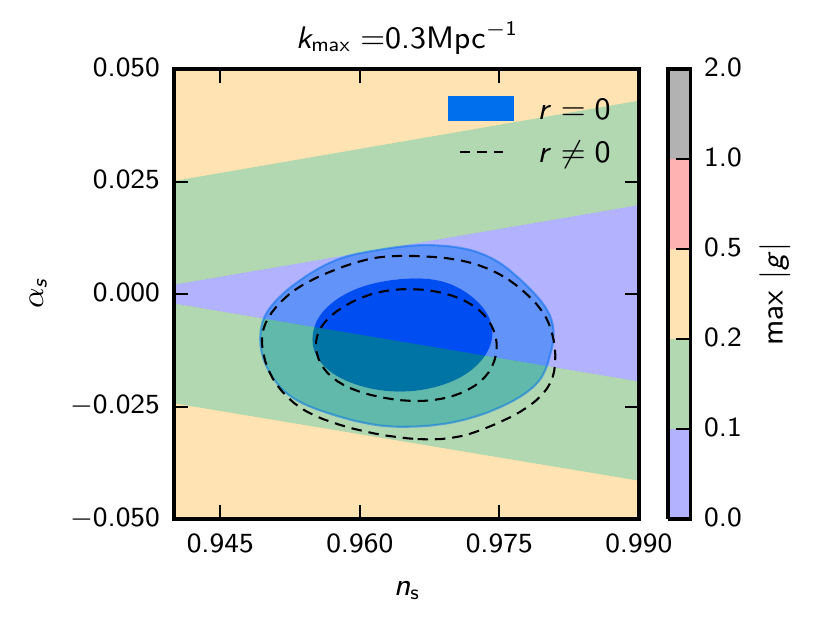}
\end{center}
\includegraphics[scale=1.0]{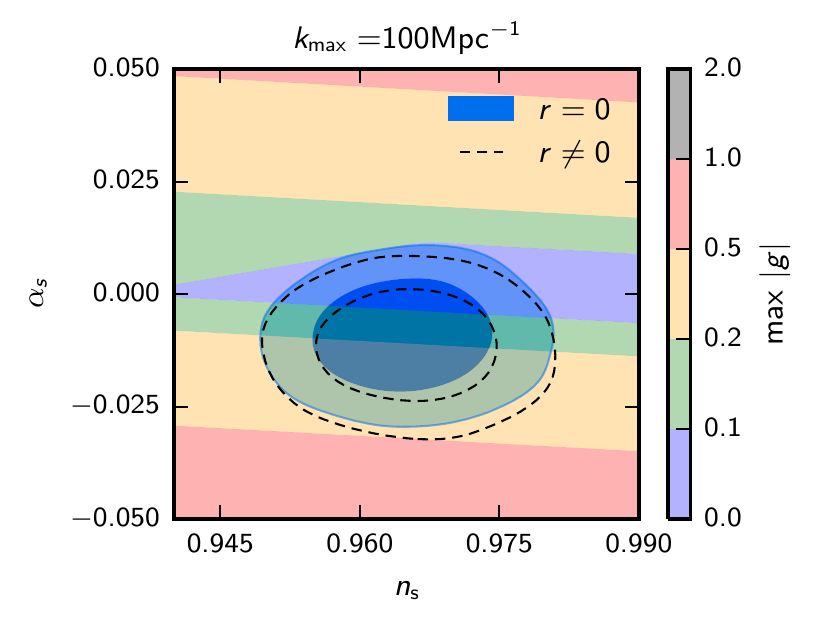}
\includegraphics[scale=1.0]{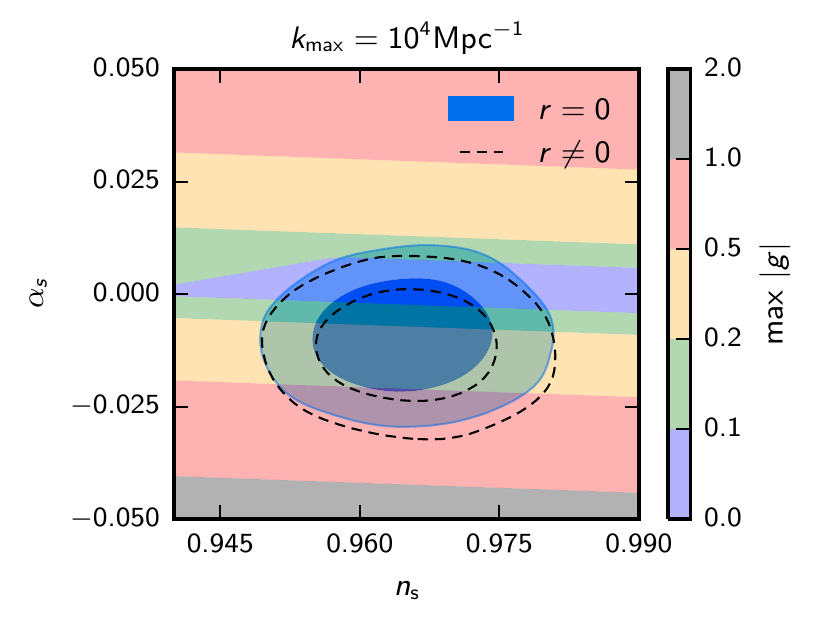}
\caption{\label{fig:alpha_fig}
Slow-roll and observational constraints on parameterizations of the power spectrum with a constant $\alpha_{s}$.
The observational contours are \emph{Planck} 2015 TT+lowTEB and joint BICEP-Keck-\emph{Planck} constraints for a constant $\alpha_{s}$ and $n_{s}$ at the pivot scale. Dashed black contours assume a null tensor-to-scalar ratio, $r$, whereas blue contours marginalise over allowed values of $r\geq 0$.
The coloured areas indicate the maximum magnitude of $g$ during the interval of time during which constrained scales
left the horizon.
Note that for $g>1$ our method breaks down as Eq.~\eqref{eq:P_f_2nd} ceases to be valid.}

\end{figure}

If we limit ourselves to the case with constant $\alpha_{s}$ (corresponding
to $N=2$ in Eq.~\eqref{eq:P_ansatz}) we have only two relevant observables:
$\alpha_{s}$ and $n_{s}$ at the pivot scale. The corresponding plots for the magnitude
of $g$ can be found in figure \ref{fig:alpha_fig}.

For the currently constrained range of scales even the 2$\sigma$ observational contours never go
beyond the $\left|g\right|<0.2$ line (which is still comfortably much less than unity). Even our
futuristic scenario with $k_{\rm max}=100\rm{Mpc^{-1}}$  has the 2$\sigma$ contour being well inside
the $\left|g\right|<0.5$ region (which corresponds to a borderline case for which the designation of ``slow-roll''
is rather dubious, but which still does not allow us to make a very strong statement\footnote{Note that
for such high values of $\left|g\right|$ we also need to worry about corrections to
Eq.~\eqref{eq:P_f_2nd} possibly becoming comparable to the observational uncertainty for the power spectrum
at the (futuristic) scale at which this maximum value is reached.}).
Only a futuristic scenario with $k_{\rm  max}=10^4\rm{Mpc^{-1}}$ would permit a measurement of constant $\alpha_s$ to
provide a strong test of slow roll. However, constraints from
spectral distortions would depend on integrals of the power spectrum over the range
$1{\rm Mpc}^{-1}\lesssim k\lesssim 10^4{\rm Mpc}^{-1}$, and cannot on their own establish
the constancy of $\alpha_s$ (even if they could accurately measure $\alpha_s$ provided it is assumed to be constant \cite{Chluba:2012we}).

\begin{figure}[h]
\begin{center}
\includegraphics[scale=1.0]{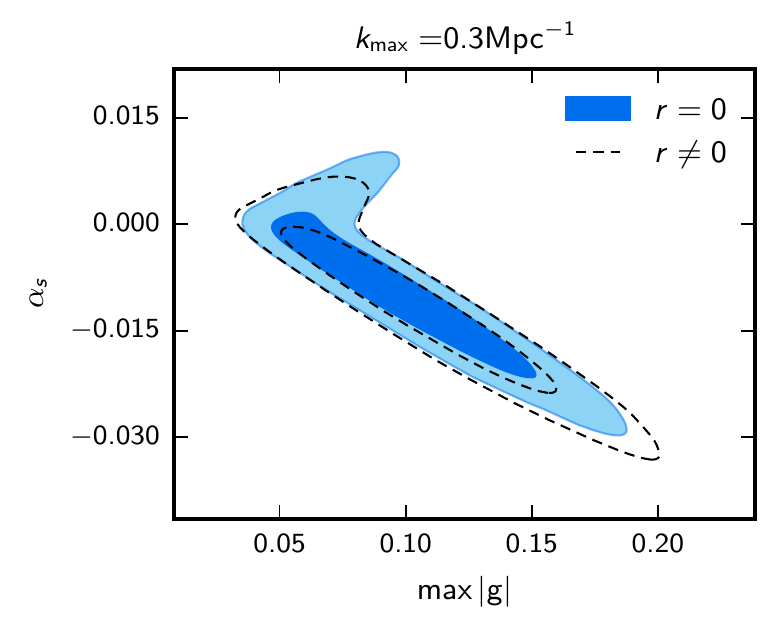}
\end{center}
\includegraphics[scale=1.0]{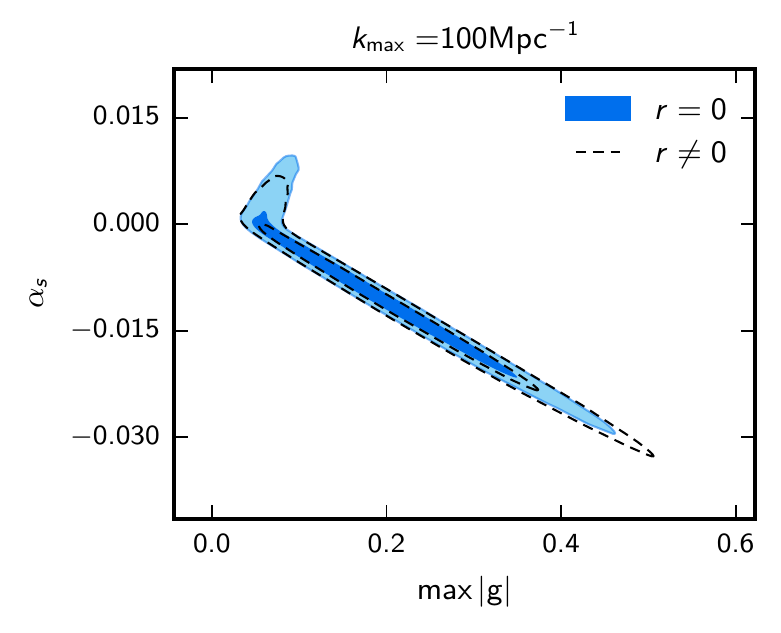}
\includegraphics[scale=1.0]{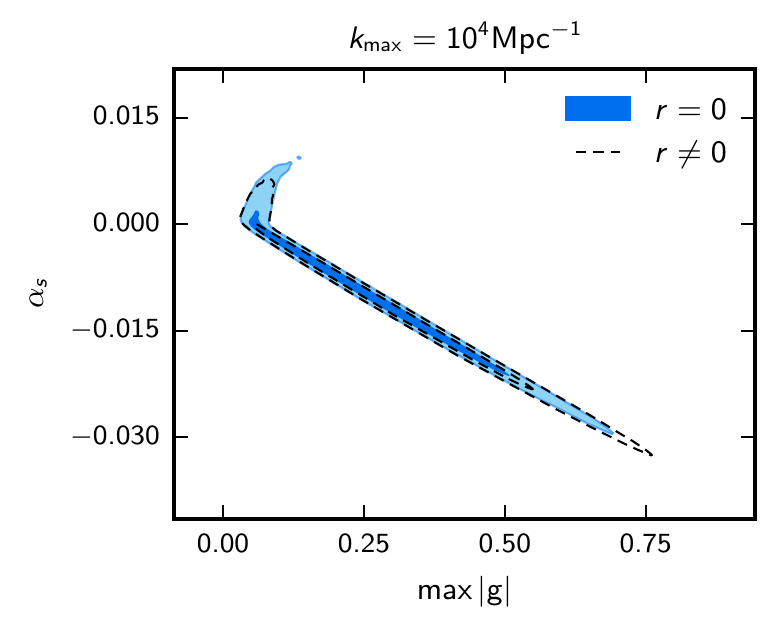}
\caption{\label{fig:galpha_fig}Bounds, over the constrained ranges of scales, on the maximum magnitude of $g$ inferred from
\emph{Planck} 2015 TT+lowTEB and joint BICEP-Keck-\emph{Planck} constraints on a constant $\alpha_{s}$ marginalized over $n_{s}$ at the pivot scale.
Filled contours assume $r=0$ whereas solid lines marginalize over allowed values of $r\geq 0$.}

\end{figure}

These conclusions are confirmed (and more easily seen) in the plots in figure \ref{fig:galpha_fig}, which show the bounds
on the maximum magnitude of $g$ inferred from the bounds on the running and the spectral index. Note that their asymmetric
boomerang shape is due to the modulus sign in $\left|g\right|$ as well as the significant deviation of the \emph{Planck} best-fit value
for $\alpha_{s}$ from zero.

\subsection{Constant running of the running ($N=3$)}

\begin{figure}[h]

\includegraphics[scale=0.95]{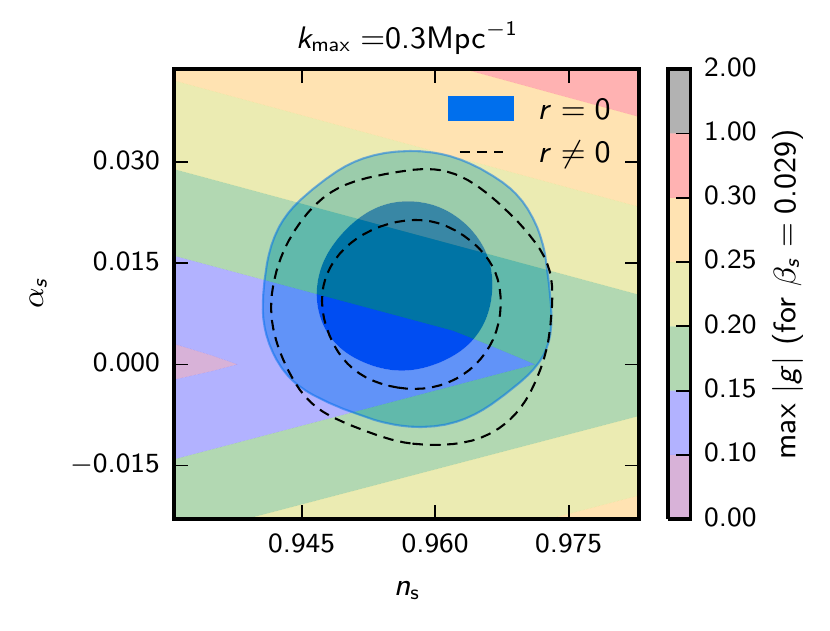}
\includegraphics[scale=0.95]{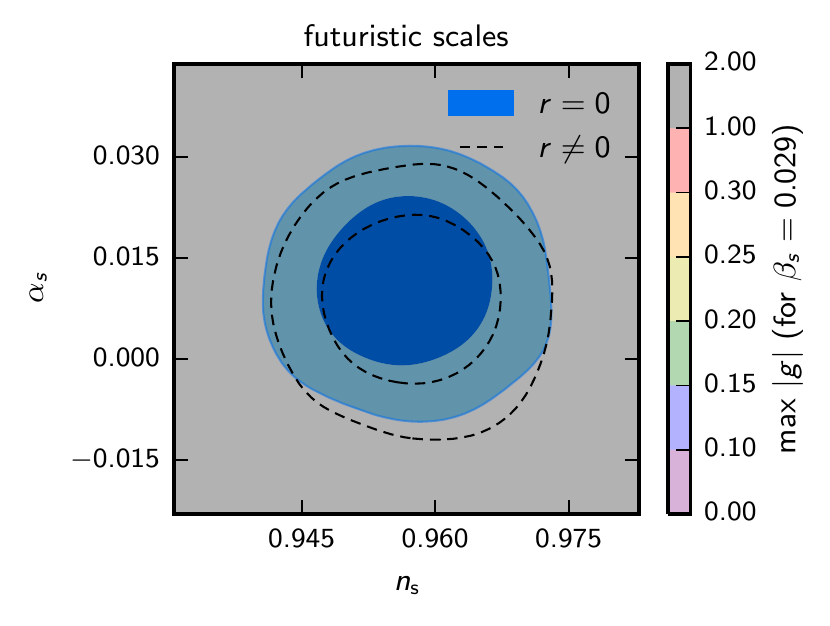}
\caption{\label{fig:beta_fig_pess}Slow-roll and observational constraints on parameterizations of the power spectrum with a constant $\beta_{s}=0.029$.
The observational contours are \emph{Planck} 2015 TT+lowTEB and joint
BICEP-Keck-\emph{Planck} constraints for $\alpha_{s}$
and $n_{s}$ at the pivot scale. Dashed black contours assume a null tensor-to-scalar ratio,
$r$,  whereas blue contours marginalise over allowed values of $r\geq 0$.
The coloured areas indicate the maximum magnitude of $g$ during the interval of time during which the constrained scales
left the horizon (``futuristic scales'' denoting both $k_{\rm max}=100\rm{Mpc^{-1}}$ and $k_{\rm max}=10^4\rm{Mpc^{-1}}$).
Note that for $g>1$
(as is the case everywhere on the plot on the right-hand side) our method breaks down as
Eq.~\eqref{eq:P_f_2nd} ceases to be valid.}
\end{figure}

If we allow the running to vary with a constant $\beta_{s}$ (corresponding to $N=3$ in Eq.~\eqref{eq:P_ansatz}) we have three relevant observables: the constant $\beta_{s}$, as well as the values of $\alpha_{s}$ and
$n_{s}$ at the pivot scale. In order to illustrate typical constraints, we present the plots
corresponding to $\beta_{s}=0.029$ (the \emph{Planck} best fit) in figure \ref{fig:beta_fig_pess} (higher values of $\beta_{s}$
would result in a more dramatic version of these plots, whereas lower values would yield
plots more similar to those in figure \ref{fig:alpha_fig}).

\begin{figure}[h]
\begin{center}
\includegraphics[scale=1.0]{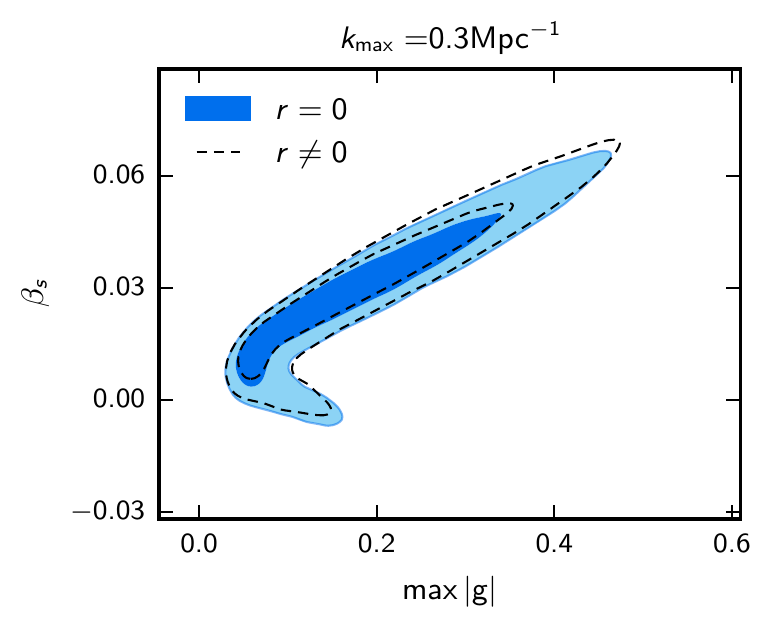}
\end{center}
\includegraphics[scale=1.0]{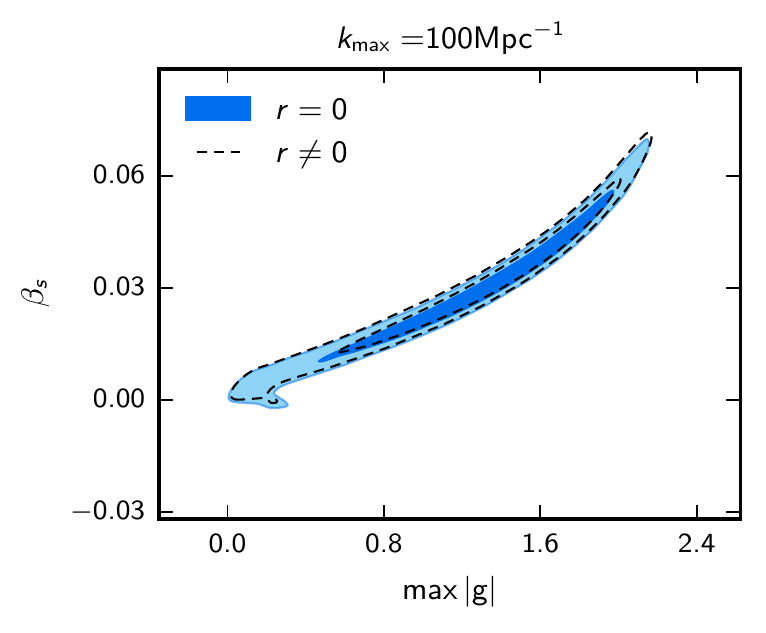}
\includegraphics[scale=1.0]{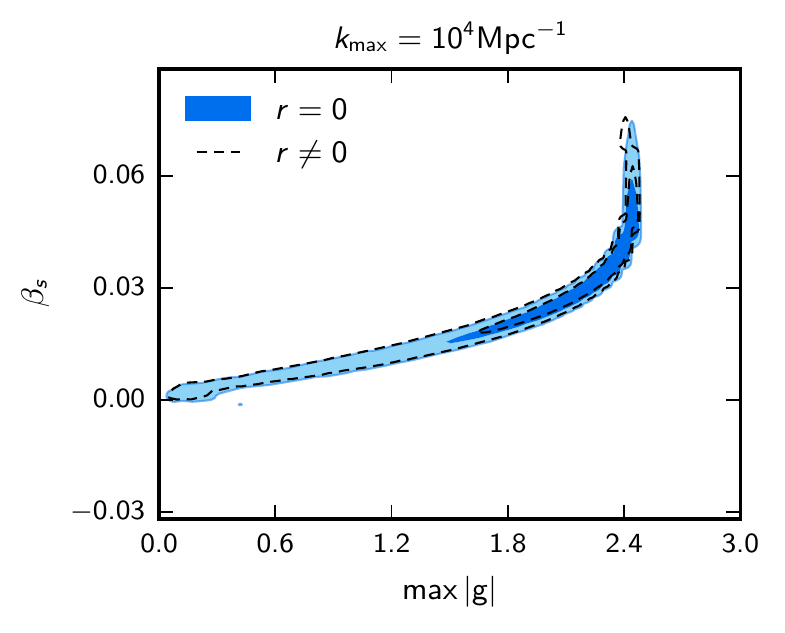}
\caption{\label{fig:beta_gfig}Bounds, over the constrained ranges of scales, on the maximum magnitude of $g$ inferred from
\emph{Planck} 2015 TT+lowTEB and BICEP-Keck constraints on a constant $\beta_{s}$ marginalized over $\alpha_{s}$ and
$n_{s}$ at the pivot scale. Filled contours assume $r=0$ whereas solid lines marginalize over allowed values of $r\geq 0$.
Note that for $g>1$ our method breaks down as Eq.~\eqref{eq:P_f_2nd} ceases to be valid.}
\end{figure}

To comment more generally on whether this class of slow-roll models can be ruled out by measuring $\beta_{s}$ over the range of its currently allowed possible values, it is easier to focus on the
 constraints on the maximum magnitude of $g$ shown in figure \ref{fig:beta_gfig}
(since they conveniently reduce the relevant three-dimensional information to simple two-dimensional contours).
The current preference for $\beta_s\ne0$ is driven by large scales, but small-scale data is consistent with constant spectral index, so as more small-scale data is added it is plausible that constraints on $\beta_s$ will converge to be closer to zero in the future.
However, if they do not, it is quite possible that a future detection of non-zero running of the running could significantly
disfavour this class of single-field slow-roll inflation, but only if information on a slightly wider range of scales is obtained
(about an extra efolding should suffice for large values of $\beta_s$ to clearly lead to high values of $\left|g\right|$, given how
some are already at the borderline $\left|g\right|\sim 0.5$.).
In particular, a future detection near \emph{Planck}'s current best fit ($\beta_{s}=0.029$)
could clearly rule out this class of slow roll.

That the larger values of constant $\beta_s$ would rule out simple slow-roll inflation models should not be a surprise.
An intuitive argument for this uses the fact that, under fairly general assumptions, to leading order in slow roll,
$n_{s}-1$ can be written in a simpler form as a sum of small $\delta_n$ parameters (of which
Eq.~\eqref{eq:ns_1st_standard} is the first-order truncation) \cite{STEWART_RUN}.
If $\beta_s = \mathcal{O}(0.05)$ and constant, $n_s$ would change by $\mathcal{O}(1)$ over the observable range
of scales, implying that this form of $n_s-1$ cannot be valid everywhere.

\subsection{Consequences for the power spectrum}

It is interesting to consider what current data say about the allowed range for
the small-scale power spectrum that could be observed by future data.
Assuming that the parameterization we have used (with constant $\beta_s$) can be extended, current Planck constraints with non-zero $\beta_s$ allow the power spectrum to grow to order unity at the smallest scales we consider (which would already be ruled out by other probes \cite{Chluba:2012we,Carr:2009jm}). Therefore, it is instructive to see how the requirement of slow roll
(as defined by a maximum
allowed magnitude of $g$ over constrained scales) would affect this extrapolation, and how that compares with the effect of the
naive expectations resulting from the imposition of the usual hierarchy on slow-roll parameters.

\begin{figure}[h]
\includegraphics[scale=1.0]{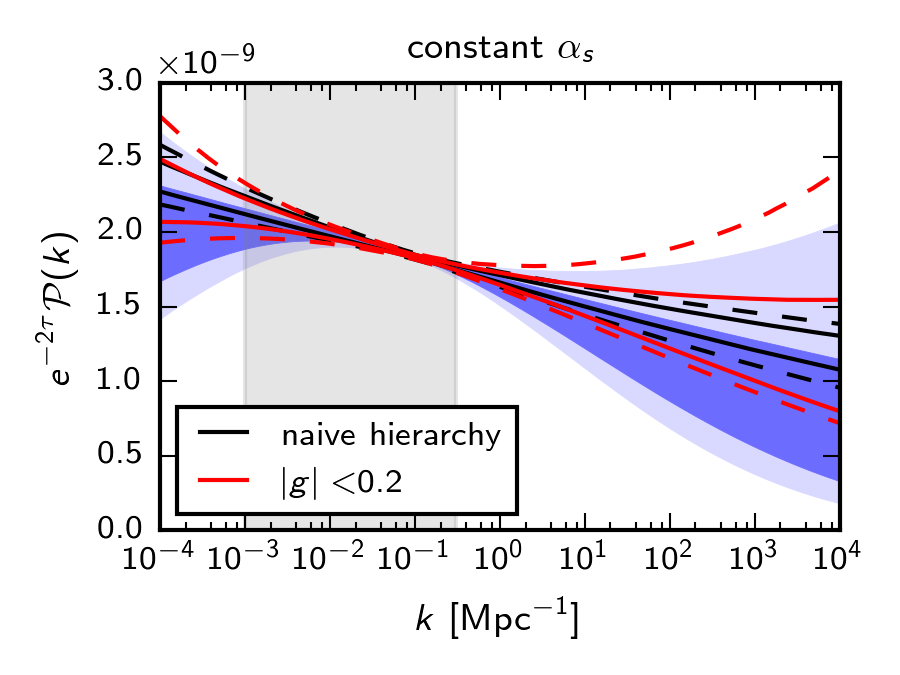}
\includegraphics[scale=1.0]{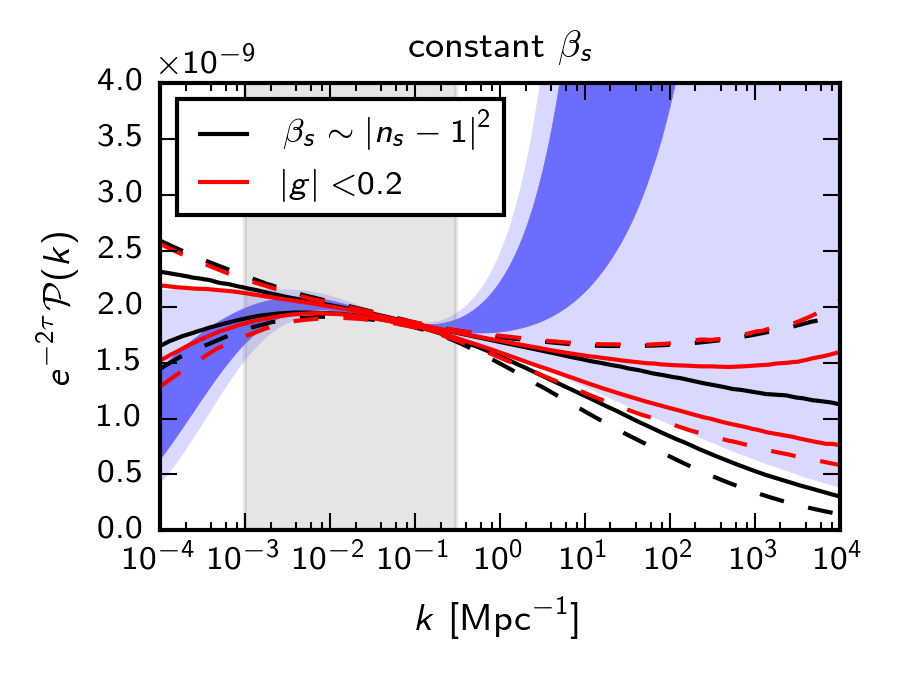}
\caption{\label{fig:Pkplots}Consequences of the imposition of slow roll (defined by the smallness of $g$) for the power spectrum
scaled by $e^{-2\tau}$, where $\tau$ is the optical depth (whose value affects the amplitude of the spectrum, but not its shape).
The blue contours represent the 68\% (dark blue) and 95\% (light blue) limits on the allowed values of the power spectrum (rescaled
by a factor of $e^{-2\tau}$) extrapolated from \emph{Planck} 2015 TT+lowTEB constraints (over gray shaded scales) assuming a
constant $\alpha_s$ (left) and a constant $\beta_{s}$ (right), for different values of $k$.
The solid and dashed red contours represent the 68\% and 95\% limits on the fraction of these spectra
for which $\left|g\right|<0.2$ for the range of scales corresponding to $10^{-3}\rm{Mpc^{-1}}<k<10^4\rm{Mpc^{-1}}$.
The solid and dashed black contours represent the 68\% and 95\% limits on the fraction of these spectra corresponding
to the unshaded regions in figure \ref{fig:motivation}
(note that for the plot on the right the limits of this region already violate the naive expectation for the magnitude of $\beta_s$).
}
\end{figure}

Our inferred constraints on the power spectrum are shown in
figure \ref{fig:Pkplots}: the assumption of slow roll leads to significantly tilted and  narrower bounds on the
small-scale power spectrum (compared to assuming only Planck constraints), especially for the case of constant $\beta_s$.

\section{Conclusions\label{sec:Conclusions}}

We devised a straightforward method to assess whether
specific observed values of the running (of the running) of the spectral index
are consistent with canonical single-field slow-roll inflation.
 We showed that slow roll is much harder to discard than simple expectations based on a hierarchy of slow-roll parameters suggest,
and in particular that for constant running any of the currently allowed values would not necessarily imply a violation of slow roll
over observable scales. However, a detection of constant $\beta_{s}$ significantly away from zero could be much more powerful\footnote{This is partly because current constraints
allow for larger constant $\beta_{s}$ than constant $\alpha_{s}$. However, mostly, it is because allowing significant higher-order
runnings implies allowing significant higher-order slow-roll parameters, which naturally makes $g$ vary faster.
In other words, $g\left(\ln\xi\right)$ computed from $f\left(\ln\xi\right)$ in Eq.~\eqref{eq:inv2_formula}
is a polynomial in $\ln\left(\xi\right)$ whose order is higher if the power spectrum has higher-order runnings.
This can also be seen from the different rates of deviation between the blue and the red limits on plots in figure \ref{fig:Pkplots}.}:
a firmer detection over currently-available scales could be enough to restrict slow-roll inflation to a region of borderline validity, and future data over a wider range of scales could invalidate slow roll for the simple parameterization of the power spectrum assumed.
%Taking into account a future increase in the range of well-constrained scales, it is possible that
%future detections may test the validity of slow-roll inflation.

% slow-roll models which go beyond the usual naive expectations for
%$\alpha_{s}$ and $\beta_{s}$ - and this turns out to be enough to show that slow roll is much harder to discard %than the
%naive expectations suggest. Furthermore, although when such models cannot be found the corresponding parameters
%cannot be completely ruled out, if future data happens to hint such regions of the observable parameter space are %important
%it is possible to improve on these results by using this method with different well-motivated parameterizations of %%the power spectrum.%

There are, however, a couple of limitations of our approach:

 - Firstly, we redefined slow-roll as meaning $\left|g\right|\ll1$, which, despite not assuming any hierarchy of slow-roll parameters,
is a somewhat stronger condition than the general definition of $\epsilon,\left|\delta_{1}\right|\ll1$
(see the conclusions of appendix \ref{sec:gappendix}). Nevertheless,
this still corresponds to a very simple and wide class of models,
including all the ones which make the wider class so popular (slow-roll formulae for the power spectrum should break down for the models left out).

 - Secondly, we follow a constructive approach: for each specific combination of
observable parameters $\{n_{s},\alpha_{s},\beta_{s},...\}$ we find a function $g\left(\ln\xi\right)$ which generates them
and check whether it breaks our weaker definition of slow-roll during the time during which observably measurable scales
crossed the horizon. Inevitably, we can only find (or fail to find) examples
of models which generate power spectra of the specific kind assumed (in the case of this work, with constant
$\alpha_{s}$  or $\beta_{s}$). In the case of constant $\alpha_s$, an existence proof is sufficient to demonstrate that models
do not violate slow roll. However, in cases where slow-roll is violated, our assumption is restrictive and different parameterizations
(e.g., involving oscillatory features, or large-scale features) might lead to different conclusions.
It would be  straightforward to generalize our method to constrain both higher-order runnings and completely
different parameterizations by making appropriate changes to Eq.~\eqref{eq:P_ansatz}.

Due to its smallness, the tensor-to-scalar ratio does not noticeably affect our results.

\section{Acknowledgements}

We thank Ewan Stewart and Daniel Passos for helpful comments.

JV is supported by an STFC studentship, CB is supported by a Royal
Society University Research Fellowship, and AL acknowledges support
from the European Research Council under the European
Union's  Seventh  Framework  Programme  (FP/2007-2013)  /
ERC Grant Agreement No.  [616170]
and from the Science and Technology Facilities Council {[}grant number ST/L000652/1{]}.

\bibliographystyle{JHEP}
\bibliography{./PhD_BIB}

\appendix

\section{$g$ and the slow-roll parameters\label{sec:gappendix}}

In this work, slow-roll is tested via the $g$ function defined in
Eq.~\eqref{eq:gdef} rather than directly via the slow-roll parameters. We relate the two here.
%It is therefore important to be able to know how the former is related
%to the latter.

\subsection{Slow-roll parameters from conformal time}

The main difficulty in relating $g$ to the slow-roll parameters stems from the terms in $g$ which are related to the conformal time.
We thus start by manipulating the usual expression for (minus) the conformal time,
\begin{equation}
\xi\left(t\right)=\intop_{t}^{\infty}\frac{dt^{'}}{a\left(t^{'}\right)}=\intop_{a\left(t\right)}^{\infty}\frac{da}{Ha^{2}}=\frac{1}{a\left(t\right)H\left(t\right)}-\intop_{a\left(t\right)}^{\infty}\frac{\dot{H}}{H^{2}}\frac{da}{a\dot{a}}=\frac{1}{a\left(t\right)H\left(t\right)}-\intop_{a\left(t\right)}^{\infty}\frac{\dot{H}}{H^{2}}\frac{da}{a^{2}H}.\label{eq:eta_(re)def}
\end{equation}
Using Eq.~\eqref{eq:eps_small}, we write it as
\begin{equation}
\xi=\frac{1}{aH}\left[1+\bar{\epsilon}\right],\label{eq:eta_epsbar}
\end{equation}
where $\bar{\epsilon}$ is given by
\begin{equation}
\bar{\epsilon}\left(\xi\right)\equiv a\left(\xi\right)H\left(\xi\right)\intop_{0}^{\xi}\epsilon(\tilde{\xi})d\tilde{\xi}=\frac{1}{\left\langle \epsilon\right\rangle _{\xi}^{-1}-1},\label{eq:eps_bar}
\end{equation}
$\left\langle \epsilon\right\rangle $ being the conformal time average
of $\epsilon$ at a given instant, defined as
\begin{equation}
\left\langle \epsilon\right\rangle _{\xi}\equiv\frac{1}{\xi}\intop_{0}^{\xi}\epsilon(\tilde{\xi})d\tilde{\xi.}\label{eq:average_eps}
\end{equation}

From Eq.~\eqref{eq:eps_small}, it can be easily seen that
\begin{equation}
\frac{d\epsilon}{d\ln\xi}=-\left(1+\bar{\epsilon}\right)\left(2\epsilon^{2}+2\epsilon\delta_{1}\right),\label{eq:deps}
\end{equation}
so variations in $\epsilon$ are second-order in slow roll and thus $\left\langle \epsilon\right\rangle $ and $\bar{\epsilon}$
are not expected to differ from $\epsilon$ at first order.

In fact, if we further assume that $\epsilon$ is well-behaved (in the sense that it can
be expressed as a Taylor series in the domain of integration of Eq.
\eqref{eq:average_eps}%
\footnote{Since the integration
domain for $\left\langle \epsilon\right\rangle $ stretches all the
way to the infinite future, a drastic departure from slow-roll at
(or even after) the end of inflation may cause this assumption to
be violated - potentially leading to $\left|\bar{\epsilon}-\epsilon\right|$ being larger
than expected. However, as long as this violation is far
enough in the future, for our purposes we can always ignore it and
pretend that slow-roll goes on forever (or alternatively stop the
integration at a very distant point before slow-roll is violated)
since we know that the curvature perturbation is conserved on very
large superhorizon scales \cite{WANDS_SEPARATE}.\label{fn:far_future}%
}), we can write $\left\langle \epsilon\right\rangle$ as
\begin{equation}
\left\langle \epsilon\right\rangle_{\xi} =\epsilon+\sum_{n=1}^{\infty}\frac{d^{n}\epsilon}{d\xi^{n}}\frac{\left(-\xi\right)^{n}}{\left(n+1\right)!}=\epsilon+\sum_{n=1}^{\infty}\left(-aH\right)^{-n}\frac{d^{n}\epsilon}{d\xi^{n}}\frac{\left(1+\bar{\epsilon}\right)^{n}}{\left(n+1\right)!}.\label{eq:epsaverageTaylor}
\end{equation}
Combining this with Eq.~\eqref{eq:deps} and its equivalent for $\delta_{n}$,
\begin{equation}
\frac{d\delta_{n}}{d\ln\xi}=-\left(1+\bar{\epsilon}\right)\left(\delta_{n+1}+n\epsilon\delta_{n}-n\delta_{1}\delta_{n}\right),\label{eq:ddeltan}
\end{equation}
it can be seen that, to second order in the slow-roll parameters,
\begin{equation}
\left\langle \epsilon\right\rangle_{\xi} \approx\epsilon+2\left(\epsilon+\sum_{p=1}^{\infty}\delta_{p}\right)\epsilon\label{eq:av_eps_final}
\end{equation}
(the right-hand side being evaluated at minus conformal time $\xi$) and
\begin{equation}
\bar{\epsilon}\approx\epsilon\left(1+3\epsilon+2\sum_{p=1}^{\infty}\delta_{p}\right).\label{eq:eps_bar_slow_final}
\end{equation}

\subsection{$g$ from $f$}

Using these results, the $f$ function defined in Eq.~\eqref{eq:fdef}
can be written as
\begin{equation}
f\left(\ln x\right)=2\pi\frac{\dot{\phi}}{H^{2}}\left[1+\bar{\epsilon}\right].\label{eq:f_epsbar}
\end{equation}
Now, using
\begin{equation}
\frac{d\ln\dot{\phi}}{d\ln\xi}=-\left(1+\bar{\epsilon}\right)\delta_{1},\label{eq:dlnphidot}
\end{equation}
\begin{equation}
\frac{d\ln H}{d\ln\xi}=\left(1+\bar{\epsilon}\right)\epsilon,\label{eq:dlnH}
\end{equation}
and
\begin{equation}
\frac{d\bar{\epsilon}}{d\ln\xi}=\left(1+\bar{\epsilon}\right)\left(\epsilon-\bar{\epsilon}+\epsilon\bar{\epsilon}\right),\label{eq:depsbar}
\end{equation}
we can find
\begin{equation}
\frac{d\ln f}{d\ln\xi}=-\bar{\epsilon}-\epsilon-\delta_{1}-\bar{\epsilon}\epsilon-\bar{\epsilon}\delta_{1}.\label{eq:dlnf}
\end{equation}
In addition, using also Eqs.~\eqref{eq:deps} and \eqref{eq:ddeltan}, we can find
\begin{equation}
\frac{d^{2}\ln f}{d\ln\xi^{2}}=\left(1+\bar{\epsilon}\right)\left(\bar{\epsilon}-\epsilon+\delta_{2}+\bar{\epsilon}\delta_{1}+\bar{\epsilon}\delta_{2}+\epsilon^{2}+2\epsilon\delta_{1}-\delta_{1}^{2}+\bar{\epsilon}\epsilon^{2}-\bar{\epsilon}\delta_{1}^{2}\right).\label{eq:d2lnf}
\end{equation}
Finally, using the definition of $g$ (Eq.~\eqref{eq:gdef}), we have
\begin{multline}
g\left(\ln x\right)=\left[
4\bar{\epsilon}+2\epsilon+3\delta_1+\delta_2+2\bar{\epsilon}^2+4\bar{\epsilon}\epsilon+5\bar{\epsilon}\delta_1
+2\bar{\epsilon}\delta_2+2\epsilon^2+4\epsilon\delta_1+2\bar{\epsilon}^2\epsilon+2\bar{\epsilon}^2\delta_1\right.\\
\left.+\bar{\epsilon}^2\delta_2
+4\bar{\epsilon}\epsilon^2+6\bar{\epsilon}\epsilon\delta_1+\bar{\epsilon}\delta_1^2+2\bar{\epsilon}^2\epsilon^2
+2\bar{\epsilon}^2\epsilon\delta_1
+\bar{\epsilon}^2\delta_1^2\right]_{\xi=\frac{x}{k}}.\label{eq:gredef}
\end{multline}
%\begin{equation}
%g\left(\ln x\right)=\left[\frac{d^{2}\ln f}{d\ln\xi^{2}}-3\frac{d\ln f}{d\ln\xi}+\left(\frac{d\ln f}{d\ln\xi}\right)^{2}\right]_{\xi=\frac{x}{k}},\label{eq:gredef}
%\end{equation}
%and hence conclude that:
%\begin{enumerate}
%\item the requirement that $g$ be very small is equivalent to requiring
%that $\bar{\epsilon}$, $\epsilon$, $\delta_{1}$, and $\delta_{2}$
%all be very small (note that if one requires $g$ to be very small
%for a wide range of scales%
%\footnote{See footnote \ref{fn:far_future}%
%} then the smallness of $\bar{\epsilon}$ follows naturally from that
%of $\epsilon$);
%\item $g=0$ should correspond to $\bar{\epsilon}=\epsilon=\delta_{1}=\delta_{2}=0$
%(or a very fine-tuned situation where, to leading order in slow-roll
%parameters, $4\bar{\epsilon}+2\epsilon+3\delta_{1}+\delta_{2}\approx0$,
%which cannot be maintained for long).
%\end{enumerate}

\section{Evaluating the integrals\label{sec:Solving-the-integrals}}

We shall see how each of the integrals in Eq.~\eqref{eq:inv2_formula}
can be calculated in a straightforward (albeit tedious) manner when
assuming Eq.~\eqref{eq:P_ansatz}.

\subsection{$\mathcal{I}_1(\xi)\equiv \intop_{0}^{\infty}\frac{dk}{k}m\left(k\xi\right)\ln\mathcal{P}\left(\ln k\right)$\label{sub:1st_int}}

Assuming Eq.~\eqref{eq:P_ansatz}, this term can be rewritten as
\begin{equation}
\mathcal{I}_1(\xi) = \sum_{n=0}^{N}\frac{\beta_{n}}{n!}\intop_{0}^{\infty}\frac{dk}{k}m\left(k\xi\right)\left(\ln\frac{k}{k_{0}}\right)^{n}\equiv\sum_{n=0}^{N}\frac{\beta_{n}}{n!}I_{n}\left(k_{0}\xi\right),\label{eq:integral1_expand}
\end{equation}
where we have defined
\begin{equation}
I_{n}\left(y\right)\equiv\intop_{0}^{\infty}\frac{dx}{x}m\left(x\right)\left(\ln\frac{x}{y}\right)^{n}=\sum_{k=0}^{n}\left(\begin{array}{c}
n\\
k
\end{array}\right)I_{n-k}\left(1\right)\left(-\ln y\right)^{k}=\sum_{k=0}^{n}\left(\begin{array}{c}
n\\
k
\end{array}\right)I_{k}\left(1\right)\left(-\ln y\right)^{n-k}.\label{eq:I_n_def}
\end{equation}

One way of iteratively computing the constant terms $I_{k}\left(1\right)$
is by considering the more general family of integrals,
\begin{equation}
\tilde{I}_{k}\left(\alpha\right)\equiv\intop_{0}^{\infty}\frac{dx}{x}m\left(x\right)\left(\ln x\right)^{k}x^{\alpha},\label{eq:Itil_def}
\end{equation}
which are related to the terms we want to compute by
\begin{equation}
\tilde{I}_{k}\left(0\right)=I_{k}\left(1\right)\label{eq:Itil_I}.
\end{equation}
The $\tilde{I}_{k}$ obey the recursive formula
\begin{equation}
\frac{\partial\tilde{I}_{k}\left(\alpha\right)}{\partial\alpha}=\tilde{I}_{k+1}\left(\alpha\right),\label{eq:Itil_recursive}
\end{equation}
which gives us a simple way to generate all the integrals we are interested
in (since we are not interested in non-integer $n$). The recursion can start from $\tilde{I}_{0}$, which can be shown to be%
\footnote{For example by first calculating the indefinite version of the corresponding
integral by expressing the trigonometric functions in Eq.~\eqref{eq:mdef}
as combinations of complex exponentials and then using the definition
of the incomplete gamma function, before taking the relevant limits
of the result.%
} the continuous version of
\begin{equation}
\tilde{I}_{0}\left(\alpha\right)=-\frac{2^{1-\alpha}}{\pi}\left(1+\alpha\right)\Gamma\left(\alpha-1\right)\sin\left(\frac{\pi}{2}\alpha\right).\label{eq:Itil_0}
\end{equation}

Putting all of this together, we can finally write the relevant integrals
up to $N=3$ as
\begin{equation}
I_{0}\left(k_{0}\xi\right)=1\label{eq:I0}
\end{equation}
\begin{equation}
I_{1}\left(k_{0}\xi\right)=-\ln\left(k_{0}\xi\right)-\gamma+2-\ln2\label{eq:I1}
\end{equation}
\begin{equation}
I_{2}\left(k_{0}\xi\right)=\ln^{2}\left(k_{0}\xi\right)+\left(-4+2\gamma+2\ln2\right)\ln\left(k_{0}\xi\right)+\frac{\pi^{2}}{12}+\gamma\left(-4+\gamma+2\ln2\right)+\left(\ln2-2\right)^{2}\label{eq:I2}
\end{equation}
\begin{multline}
I_{3}\left(k_{0}\xi\right)=-\ln^{3}\left(k_{0}\xi\right)+3\left(2-\gamma-\ln2\right)\ln^{2}\left(k_{0}\xi\right)\\-\frac{1}{4}\left(12\left(\gamma-2\right)^{2}+\pi^{2}+12\ln2\left(-4+2\gamma+\ln2\right)\right)\ln\left(k_{0}\xi\right)\\
+\frac{1}{4}\left(48+2\pi^{2}-8\zeta\left(3\right)-\gamma\left(48+4\left(\gamma-6\right)\gamma+\pi^{2}\right)^{^{^{}}}\right.\\
\left.-4\ln^{3}2+24\ln^{2}2-12\gamma\ln^{2}2-\left(12\left(\gamma-2\right)^{2}+\pi^{2}\right)\ln2\right),\label{eq:I3}
\end{multline}
where $\gamma\simeq0.5772$ is the Euler-Mascheroni constant and $\zeta$
is the Riemann zeta function, the next integral (which it turns out
will be relevant further ahead) being given by
\begin{multline}
I_{4}\left(k_{0}\xi\right)=\ln^{4}\left(k_{0}\xi\right)+4\left(-2+\gamma+\ln2\right)\ln^{3}\left(k_{0}\xi\right)\\
+\frac{1}{2}\left(12\gamma^{2}+\pi^{2}+24\gamma\left(\ln2-2\right)+12\left(\ln2-2\right)^{2}\right)\ln^{2}\left(k_{0}\xi\right)\\
+\left(8\zeta\left(3\right)-48+4\ln^{3}2-24\ln^{2}2+48\ln2+4\gamma^{3} {}^{^{^{^{}}}}\right.\\
\left.
+\gamma\left(\pi^{2}+12\left(\ln2-2\right)^{2}\right)+12\gamma^{2}\left(\ln2-2\right)+\pi^{2}\left(\ln2-2\right)\right)\times\ln\left(k_{0}\xi\right)\\
-16\zeta\left(3\right)+\gamma\left(4\left(2\zeta\left(3\right)-12+\ln^{3}2-6\ln^{2}2+12\ln2\right)+\pi^{2}\left(\ln2-2\right)\right)\\
+8\zeta\left(3\right)\ln2+\frac{19\pi^{4}}{240}+2\pi^{2}+\gamma^{4}+48+\ln^{4}2-8\ln^{3}2\\
+\frac{1}{2}\pi^{2}\ln^{2}2+24\ln^{2}2-2\pi^{2}\ln2-48\ln2+4\gamma^{3}\left(\ln2-2\right)+\frac{1}{2}\gamma^{2}\left(\pi^{2}+12\left(\ln2-2\right)^{2}\right).\label{eq:I4}
\end{multline}

\subsection{\textmd{\normalsize{
\newform{
$\mathcal{I}_2(\xi)\equiv -\frac{\pi^2}{8}\intop_{0}^{\infty}\frac{dk}{k}m\left(k\xi\right)
\left[\intop_{0}^{\infty}\frac{dl}{l}
 \frac{\mathcal{P}'(\ln l)}{\mathcal{P}(\ln l)}
 \intop_{0}^{\infty}\frac{d\zeta}{\zeta}m\left(k\zeta\right)m\left(l\zeta\right)
 \right]^{2}$
}{
$\mathcal{I}_2(\xi)\equiv -\frac{1}{2\pi^{2}}\intop_{0}^{\infty}\frac{dk}{k}m\left(k\xi\right)\left[\intop_{0}^{\infty}\frac{dl}{l}\ln\left|\frac{k+l}{k-l}\right|\frac{\mathcal{P}^{\prime}\left(\ln l\right)}{\mathcal{P}\left(\ln l\right)}\right]^{2}$
}}}}

Assuming Eq.~\eqref{eq:P_ansatz}, this term can be written as
\newform{
\begin{equation}
\mathcal{I}_2(\xi) = -\frac{\pi^2}{8}\intop_{0}^{\infty}\frac{dk}{k}m\left(k\xi\right)
\left[
\sum_{n=0}^{N-1}\frac{\beta_{n+1}}{n!}
\intop_{0}^{\infty}\frac{dl}{l}
 \intop_{0}^{\infty}\frac{d\zeta}{\zeta}m\left(k\zeta\right)m\left(l\zeta\right)
 \left(\ln\frac{l}{k_{0}}\right)^{n}
 \right]^{2}.\label{eq:integral2_expand}
\end{equation}
}{
\begin{equation}
-\frac{1}{2\pi^{2}}\intop_{0}^{\infty}\frac{dk}{k}m\left(k\xi\right)\left[\sum_{n=0}^{N-1}\frac{\beta_{n+1}}{n!}\intop_{0}^{\infty}\frac{dl}{l}\ln\left|\frac{k+l}{k-l}\right|\left(\ln\frac{l}{k_{0}}\right)^{n}\right]^{2}.\label{eq:integral2_expand}
\end{equation}
}
\newform{
It is convenient to focus first on the integral being squared, which we can write as a sum of terms of the form
\begin{equation}
\intop_{0}^{\infty}\frac{dl}{l}\intop_{0}^{\infty}\frac{d\zeta}{\zeta}m\left(k\zeta\right)m\left(l\zeta\right)\left(\ln\frac{l}{k_{0}}\right)^{n}
=\intop_{0}^{\infty}\frac{d\zeta}{\zeta}m\left(k\zeta\right)I_{n}\left(k_{0}\zeta\right),\label{eq:logabs_I}
\end{equation}
}{
It is convenient to focus first on the integral being squared.
Using
the known relation \cite{Stewart_inverse2}
\begin{equation}
\intop_{0}^{\infty}\frac{d\zeta}{\zeta}m\left(k\zeta\right)m\left(l\zeta\right)=\frac{2}{\pi^{2}}\ln\left|\frac{k+l}{k-l}\right|\label{eq:logabs_integral}
\end{equation}
we can write
\begin{equation}
\intop_{0}^{\infty}\frac{dl}{l}\ln\left|\frac{k+l}{k-l}\right|\left(\ln\frac{l}{k_{0}}\right)^{n}
=
\frac{\pi^{2}}{2}\intop_{0}^{\infty}\frac{dl}{l}\intop_{0}^{\infty}\frac{d\zeta}{\zeta}m\left(k\zeta\right)m\left(l\zeta\right)\left(\ln\frac{l}{k_{0}}\right)^{n}=\frac{\pi^{2}}{2}\intop_{0}^{\infty}\frac{d\zeta}{\zeta}m\left(k\zeta\right)I_{n}\left(k_{0}\zeta\right),\label{eq:logabs_I}
\end{equation}
}
where we changed the order of integration and
used the definition of $I_{n}$ from Eq.~\eqref{eq:I_n_def}. Given
that these functions can be quite messy in appearance, but are always
polynomials in $\ln\left(k_{0}\xi\right)$, we write them as
\begin{equation}
I_{n}\left(k_{0}\xi\right)\equiv\sum_{i=0}^{n}c_{n}\left[i\right]\ln^{i}\left(k_{0}\xi\right),\label{eq:cn_def}
\end{equation}
where the $c_{n}\left[i\right]$ coefficients can be found as described in subsection \ref{sub:1st_int} (the relevant ones
being trivially obtained by comparison with Eqs.
\eqref{eq:I0}, \eqref{eq:I1}, \eqref{eq:I2}, \eqref{eq:I3}, and \eqref{eq:I4}).
This ``inner'' integral thus becomes
\newform{
\begin{equation}
\intop_{0}^{\infty}\frac{d\zeta}{\zeta}m\left(k\zeta\right)I_{n}\left(k_{0}\zeta\right)=
\sum_{i=0}^{n}c_{n}\left[i\right]\intop_{0}^{\infty}\frac{d\zeta}{\zeta}m\left(k\zeta\right)\ln^{i}\left(k_{0}\zeta\right)=\sum_{i=0}^{n}\sum_{j=0}^{i}c_{n}\left[i\right]c_{i}\left[j\right]\left(\ln\frac{k}{k_{0}}\right)^{j}\label{eq:logabscn2}.
\end{equation}
Substituting this into Eq.~\eqref{eq:integral2_expand} and changing the order of summation  we are left with
\begin{equation}
\sum_{n=0}^{N-1}\frac{\beta_{n+1}}{n!}
\intop_{0}^{\infty}\frac{dl}{l}
 \intop_{0}^{\infty}\frac{d\zeta}{\zeta}m\left(k\zeta\right)m\left(l\zeta\right)
 \left(\ln\frac{l}{k_{0}}\right)^{n}
=\sum_{j=0}^{N-1}\tilde{c}_{N}\left[j\right]\left(\ln\frac{k}{k_{0}}\right)^{j},\label{eq:inner_int}
\end{equation}
where we have defined
\begin{equation}
\tilde{c}_{N}\left[j\right]\equiv\sum_{n=j}^{N-1}\sum_{i=j}^{n}\frac{\beta_{n+1}}{n!}c_{n}\left[i\right]c_{i}\left[j\right].\label{eq:cNtil_def}
\end{equation}
Finally, we can tackle the full double integral, writing
\begin{equation}
\mathcal{I}_2(\xi) = -\frac{\pi^2}{8}\sum_{i=0}^{N-1}\sum_{j=0}^{N-1}\tilde{c}_{N}\left[i\right]\tilde{c}_{N}\left[j\right]\intop_{0}^{\infty}\frac{dk}{k}m\left(k\xi\right)\left(\ln\frac{k}{k_{0}}\right)^{i+j}\label{eq:outer_int_expand}
\end{equation}
which, using Eq.~\eqref{eq:cn_def} once more, simplifies to
\begin{equation}
\mathcal{I}_2(\xi) = -\frac{\pi^2}{8}\sum_{i=0}^{N-1}\sum_{j=0}^{N-1}\sum_{s=0}^{i+j}\tilde{c}_{N}\left[i\right]\tilde{c}_{N}\left[j\right]c_{i+j}\left[s\right]\ln^{s}\left(k_{0}\xi\right).\label{eq:outer_int}
\end{equation}
Here, we finally see why Eq.~\eqref{eq:I4} was needed (since $s$ can
vary up to $s=4$ for $N=3$).

}{
\begin{equation}
\intop_{0}^{\infty}\frac{dl}{l}\ln\left|\frac{k+l}{k-l}\right|\left(\ln\frac{l}{k_{0}}\right)^{n}=\frac{\pi^{2}}{2}\sum_{i=0}^{n}c_{n}\left[i\right]\intop_{0}^{\infty}\frac{d\zeta}{\zeta}m\left(k\zeta\right)\ln^{i}\left(k_{0}\zeta\right)=\frac{\pi^{2}}{2}\sum_{i=0}^{n}\sum_{j=0}^{i}c_{n}\left[i\right]c_{i}\left[j\right]\left(\ln\frac{k}{k_{0}}\right)^{j}\label{eq:logabscn2}
\end{equation}
and, substituting this into Eq.~\eqref{eq:integral2_expand} and changing
the order of summation, we are left with
\begin{equation}
\intop_{0}^{\infty}\frac{dl}{l}\ln\left|\frac{k+l}{k-l}\right|\frac{\mathcal{P}^{\prime}\left(\ln l\right)}{\mathcal{P}\left(\ln l\right)}=\sum_{j=0}^{N-1}\tilde{c}_{N}\left[j\right]\left(\ln\frac{k}{k_{0}}\right)^{j},\label{eq:inner_int}
\end{equation}
where we have defined
\begin{equation}
\tilde{c}_{N}\left[j\right]\equiv\sum_{n=j}^{N-1}\sum_{i=j}^{n}\frac{\pi^{2}}{2}\frac{\beta_{n+1}}{n!}c_{n}\left[i\right]c_{i}\left[j\right].\label{eq:cNtil_def}
\end{equation}

At last, we can tackle the full double integral, writing
\begin{equation}
-\frac{1}{2\pi^{2}}\intop_{0}^{\infty}\frac{dk}{k}m\left(k\xi\right)\left[\intop_{0}^{\infty}\frac{dl}{l}\ln\left|\frac{k+l}{k-l}\right|\frac{\mathcal{P}^{\prime}\left(\ln l\right)}{\mathcal{P}\left(\ln l\right)}\right]^{2}=\sum_{i=0}^{N-1}\sum_{j=0}^{N-1}-\frac{1}{2\pi^{2}}\tilde{c}_{N}\left[i\right]\tilde{c}_{N}\left[j\right]\intop_{0}^{\infty}\frac{dk}{k}m\left(k\xi\right)\left(\ln\frac{k}{k_{0}}\right)^{i+j}\label{eq:outer_int_expand}
\end{equation}
which, using Eq.~\eqref{eq:cn_def} once more, simplifies to
\begin{equation}
-\frac{1}{2\pi^{2}}\intop_{0}^{\infty}\frac{dk}{k}m\left(k\xi\right)\left[\intop_{0}^{\infty}\frac{dl}{l}\ln\left|\frac{k+l}{k-l}\right|\frac{\mathcal{P}^{\prime}\left(\ln l\right)}{\mathcal{P}\left(\ln l\right)}\right]^{2}=\sum_{i=0}^{N-1}\sum_{j=0}^{N-1}\sum_{s=0}^{i+j}-\frac{1}{2\pi^{2}}\tilde{c}_{N}\left[i\right]\tilde{c}_{N}\left[j\right]c_{i+j}\left[s\right]\ln^{s}\left(k_{0}\xi\right),\label{eq:outer_int}
\end{equation}
where we finally see why Eq.~\eqref{eq:I4} was needed (since $s$ can
vary up to $s=4$ for $N=3$).
}

\subsection{
\newform{
$
\mathcal{I}_3(\xi) \equiv
\frac{\pi}{2}\intop_{0}^{\infty}\frac{dl}{l} \frac{\mathcal{P}'(\ln l)}{\mathcal{P}(\ln l)} \intop_{0}^{\infty}\frac{dq}{q}
\frac{\mathcal{P}'(\ln q)}{\mathcal{P}(\ln q)}
\intop_{0}^{\infty}\frac{d\zeta}{\zeta}m\left(l\zeta\right)\intop_{0}^{\infty}\frac{dk}{k^{2}}m\left(k\xi\right)m\left(k\zeta\right)\intop_{\zeta}^{\infty}\frac{d\chi}{\chi^{2}}m\left(q\chi\right)
$}{
$\mathcal{I}_3(\xi) \equiv \intop_{0}^{\infty}\frac{dl}{l}\intop_{0}^{\infty}\frac{dq}{q}M\left(l\xi,q\xi\right)\frac{\mathcal{P}^{\prime}\left(\ln l\right)}{\mathcal{P}\left(\ln l\right)}\frac{\mathcal{P}^{\prime}\left(\ln q\right)}{\mathcal{P}\left(\ln q\right)}$
}}

Assuming Eq.~\eqref{eq:P_ansatz}, this term can be written as
\newform{
}{
\begin{equation}
\sum_{n=0}^{N-1}\sum_{s=0}^{N-1}\frac{\beta_{n+1}}{n!}\frac{\beta_{s+1}}{s!}\intop_{0}^{\infty}\frac{dl}{l}\intop_{0}^{\infty}\frac{dq}{q}M\left(l\xi,q\xi\right)\left(\ln\frac{l}{k_{0}}\right)^{n}\left(\ln\frac{q}{k_{0}}\right)^{s}\label{eq:integral3_expand}
\end{equation}
and we can use the known relation \cite{Stewart_inverse2}
\begin{equation}
\intop_{0}^{\infty}\frac{d\zeta}{\zeta}m\left(l\zeta\right)\intop_{0}^{\infty}\frac{dk}{k^{2}}m\left(k\xi\right)m\left(k\zeta\right)\intop_{\zeta}^{\infty}\frac{d\chi}{\chi^{2}}m\left(q\chi\right)=\frac{2}{\pi}M\left(l\xi,q\xi\right)\label{eq:M_integral}
\end{equation}
to express it as
}
\begin{multline}
\mathcal{I}_3(\xi) =
\frac{\pi}{2}\sum_{n=0}^{N-1}\sum_{s=0}^{N-1}\frac{\beta_{n+1}}{n!}\frac{\beta_{s+1}}{s!}
\times \\ \intop_{0}^{\infty}\frac{dl}{l}\intop_{0}^{\infty}\frac{dq}{q}\intop_{0}^{\infty}\frac{d\zeta}{\zeta}m\left(l\zeta\right)\intop_{0}^{\infty}\frac{dk}{k^{2}}m\left(k\xi\right)m\left(k\zeta\right)\intop_{\zeta}^{\infty}\frac{d\chi}{\chi^{2}}m\left(q\chi\right)\left(\ln\frac{l}{k_{0}}\right)^{n}\left(\ln\frac{q}{k_{0}}\right)^{s},\label{eq:2sum_5int}
\end{multline}
which we now can solve using a similar method to the previous subsections. For example, using Eq.~\eqref{eq:cn_def}
and integrating with respect to $l$ and $q$ we are left with
\begin{equation}
\sum_{n=0}^{N-1}\sum_{s=0}^{N-1}\sum_{i=0}^{n}\sum_{j=0}^{s}\frac{\pi}{2}\frac{\beta_{n+1}}{n!}\frac{\beta_{s+1}}{s!}c_{n}\left[i\right]c_{s}\left[j\right]\intop_{0}^{\infty}\frac{d\zeta}{\zeta}\ln^{i}\left(k_{0}\zeta\right)\intop_{0}^{\infty}\frac{dk}{k^{2}}m\left(k\xi\right)m\left(k\zeta\right)\intop_{\zeta}^{\infty}\frac{d\chi}{\chi^{2}}\ln^{j}\left(k_{0}\chi\right).\label{eq:2sum_5int-1}
\end{equation}
Focussing on the integral with respect to $\chi$, a simple
change of variables gives
\begin{equation}
\intop_{\zeta}^{\infty}\frac{d\chi}{\chi^{2}}\ln^{j}\left(k_{0}\chi\right)=k_{0}\intop_{k_{0}\zeta}^{\infty}\frac{dx}{x^{2}}\ln^{j}\left(x\right)=k_{0}\intop_{\ln\left(k_{0}\zeta\right)}^{\infty}t^{j}e^{-t}dt\equiv k_{0}\Gamma\left(j+1,\ln\left(k_{0}\zeta\right)\right)\label{eq:incomplete_gamma}
\end{equation}
where $\Gamma$ is the (upper) incomplete gamma function. Since $j$
is an integer, this can be explicitly written as the type of polynomial
we are interested in by using the known relation
\begin{equation}
\Gamma\left(j+1,\ln\left(k_{0}\zeta\right)\right)=\frac{j!}{k_{0}\zeta}\sum_{\sigma=0}^{j}\frac{\ln^{\sigma}\left(k_{0}\zeta\right)}{\sigma!}.\label{eq:integer_inc_gamma}
\end{equation}
The full integral thus becomes
\begin{equation}
\sum_{n=0}^{N-1}\sum_{s=0}^{N-1}\sum_{i=0}^{n}\sum_{j=0}^{s}\sum_{\sigma=0}^{j}\frac{\beta_{n+1}}{n!}\frac{\beta_{s+1}}{s!}\frac{\pi}{2}\frac{j!}{\sigma!}c_{n}\left[i\right]c_{s}\left[j\right]\intop_{0}^{\infty}\frac{d\zeta}{\zeta^{2}}\intop_{0}^{\infty}\frac{dk}{k^{2}}m\left(k\xi\right)m\left(k\zeta\right)\ln^{i+\sigma}\left(k_{0}\zeta\right),\label{eq:2sum_5int-1-1}
\end{equation}
which can be tackled by noticing that
\begin{multline}
\intop_{0}^{\infty}\frac{d\zeta}{\zeta^{2}}m\left(k\zeta\right)\ln^{i+\sigma}\left(k_{0}\zeta\right)=k\intop_{0}^{\infty}\frac{dx}{x^{2}}m\left(x\right)\ln^{i+\sigma}\left(\frac{k_{0}}{k}x\right)\\
=\sum_{\rho=0}^{i+\sigma}\binom{i+\sigma}{\rho}
k\left(\ln\frac{k_{0}}{k}\right)^{\rho}\intop_{0}^{\infty}\frac{dx}{x^{2}}m\left(x\right)\left(\ln x\right)^{i+\sigma-\rho}\equiv\sum_{\rho=0}^{i+\sigma}
\binom{i+\sigma}{\rho}
k\left(\ln\frac{k_{0}}{k}\right)^{\rho}\tilde{I}_{i+\sigma-\rho}\left(-1\right),\label{eq:Itil-1}
\end{multline}
where in the last step we used the definition of $\tilde{I}$
from Eq.~\eqref{eq:Itil_def}. The full integral is therefore reduced
to a sextuple sum of single integrals,
\begin{equation}
\sum_{n=0}^{N-1}\sum_{s=0}^{N-1}\sum_{i=0}^{n}\sum_{j=0}^{s}\sum_{\sigma=0}^{j}\sum_{\rho=0}^{i+\sigma}\frac{\beta_{n+1}}{n!}\frac{\beta_{s+1}}{s!}\frac{\pi}{2}
%\left(\begin{array}{c}i+\sigma\\\rho\end{array}\right)
\binom{i+\sigma}{\rho}
\frac{j!}{\sigma!}\tilde{I}_{i+\sigma-\rho}\left(-1\right)c_{n}\left[i\right]c_{s}\left[j\right]\intop_{0}^{\infty}\frac{dk}{k}m\left(k\xi\right)\left(-\ln\frac{k}{k_{0}}\right)^{\rho}.\label{eq:12go}
\end{equation}
The remaining integral is simply $\left(-1\right)^{\rho}I_{\rho}\left(k_{0}\xi\right)$,
allowing us to write the final result as the following septuple sum
of known and given terms (keeping in mind that the method for finding
out any $\tilde{I}$ was shown in subsection \ref{sub:1st_int})
\begin{equation}
\sum_{n=0}^{N-1}\sum_{s=0}^{N-1}\sum_{i=0}^{n}\sum_{j=0}^{s}\sum_{\sigma=0}^{j}\sum_{\rho=0}^{i+\sigma}\sum_{\delta=0}^{\rho}\frac{\beta_{n+1}}{n!}\frac{\beta_{s+1}}{s!}\frac{\pi}{2}
%\left(\begin{array}{c}i+\sigma\\\rho\end{array}\right)
\binom{i+\sigma}{\rho}\frac{j!}{\sigma!}\left(-1\right)^{\rho}\tilde{I}_{i+\sigma-\rho}\left(-1\right)c_{n}\left[i\right]c_{s}\left[j\right]c_{\rho}\left[\delta\right]\ln^{\delta}\left(k_{0}\xi\right),\label{eq:int3_final}
\end{equation}

\end{document}